\documentstyle[aps,preprint,tighten,eqsecnum,floats,epsf,epsfig,rotate,prd]
{revtex}

\newcommand{\beq}{\begin{equation}}
\newcommand{\eeq}{\end{equation}}
\newcommand{\beqa}{\begin{eqnarray}}
\newcommand{\eeqa}{\end{eqnarray}}

\newcommand{\be}{\begin{eqnarray}}
\newcommand{\ee}{\end{eqnarray}}
\newcommand{\sbe}{\begin{eqnarray*}}
\newcommand{\see}{\end{eqnarray*}}

\newbox\rotbox

\begin{document}
\preprint{Preprint Number: \parbox[t]{50mm}{ADP-99-43/T380}}
\draft
%_______________________ Title, Authors ____________________________________
\title{On conditions for the nonperturbative equivalence of \\
         ultraviolet cut-off and dimensional regularization schemes}

\author{
        Ay{\c s}e K{\i}z{\i}lers{\" u}$^a$
            \footnote{E-mail:~akiziler@physics.adelaide.edu.au},
        Tom Sizer$^a$
            \footnote{E-mail:~tsizer@physics.adelaide.edu.au}
          and
        Anthony G.\ Williams$^{a,b}$
            \footnote{E-mail:~awilliam@physics.adelaide.edu.au}
        \vspace*{2mm}\\}

\address{
   $^a$ Special Research Centre for the Subatomic Structure of Matter and \\
   Department of Physics and Mathematical Physics, \\
   University of Adelaide, 5005, Australia
   \vspace*{2mm}\\
   $^b$ Department of Physics and SCRI,
   Florida State University, \\
   Tallahassee, Florida 32306-3016
}
%\date{}
%
\maketitle
%%-------------------------------------------------------------------%
%
\begin{abstract}

  We consider procedures through which an ultraviolet cut-off
  regularization scheme can be modified to reproduce the same
  results for
  nonperturbative renormalized Green's functions as obtained
  from a dimensional regularization scheme.  These issues are
  considered within the Dyson-Schwinger equation framework,
  where ultraviolet cut-off regularization
  can lead to explicit violations of gauge invariance.  As a
  specific illustration, we consider the electron
  self-energy in quenched QED$_4$ in both schemes and establish
  those procedures for which the UV cut-off scheme can be expected to
  lead to the dimensional regularization results.
  We also compare results from precise numerical studies using the two
  types of regularization.

\end{abstract}

\newpage
\section{Introduction}
\label{sec_intro}

In order to study quantum field theories in the nonperturbative
regime it is essential to have appropriate regularization
schemes which respect the symmetries of the underlying theory.
In the lattice approach\cite{Rothe:1992,Montvay:1994,Gupta:1998}
to nonperturbative studies of gauge theories the gauge fields are
represented by links and the action is formulated
in terms of these links in order to maintain gauge invariance
by construction.  This is the case even though
the finite lattice spacing is acting as a form of ultraviolet
regulator.  It is useful to augment lattice studies by other
nonperturbative methods such as studies of Dyson-Schwinger
equations (DSE)\cite{DSE_Review,MiranskReview,FGMS}.
Studies using the DSE must always involve some form of truncation
of the infinite tower of coupled integral equations and hence can
never be a first principles approach.
Nevertheless, DSE are a useful complement to lattice studies
and through their self-consistent nature can provide
important insights into the nonperturbative behavior of quantum field
theories.  In addition, information obtained from the lattice
can place significant constraints on the DSE approach, which can
further enhance its usefulness.

One difficulty facing DSE integral equation studies is that the
use of an ultraviolet (UV) cut-off to regulate
the integrations will in general lead to an
explicit violation of gauge invariance.  In other words, the
renormalized Green's functions calculated within a DSE study using
a UV cut-off will in general contain unacceptable explicit
gauge-invariance violating contributions, unless specific
steps are taken to remove them. On the other hand, DSE studies
implemented using a gauge--invariant regularization scheme such
as dimensional regularization will have no such undesirable
explicit gauge-invariance violating contributions.

Only recently have explicit numerical DSE studies been
succesfully performed using dimensional
regularization\cite{Schreiber:1998,Gusynin:1999}.
The subject of these initial studies was quenched QED$_4$,
which, while not a physically realistic theory, has
the advantage of being simple enough that it
is an excellent testing ground for nonperturbative techniques
such as DSE and lattice studies.
Renormalized quantities calculated within a dimensional
regularization scheme can be compared directly with those
obtained using a UV cut-off scheme, (provided of course that
the same renormalization conditions are imposed).
This is just what was done for studies of the fermion
propagator and dynamical chiral symmetry breaking using
dimensional regularization in
Refs.~\cite{Schreiber:1998,Gusynin:1999}, where direct
comparisons of results for the fermion propagator and
critical coupling $\alpha_c$ were made with results obtained
using ultraviolet cut-off regularization
\cite{qed4_hw,qed4_hrw,qed4_hsw}.
It was found that, with an appropriate modification to the naive
UV cut-off treatment and within the currently achieved numerical
precision, the results were the same as those obtained from
the more computationally demanding dimensional regularization
approach.  In nonperturbative studies the UV cut-off regularization
scheme can
have significant computational advantages over the dimensional
regularization scheme, where a careful extrapolation to
the $\epsilon\to 0$ limit must be taken numerically.

The purpose of the present work is to exploit this recent
development of nonperturbative dimensional regularization
in order to help motivate and establish general principles for
removing the unwanted explicit gauge-violating contributions in the UV
cut-off regularization approach. This is to be achieved by
 imposing translational invariance on to  cut-off
regularization. In order to achieve this we are free to add terms
which would vanish in any translationally invariant scheme. Hence
one must choose an arbitrary centre for the 4-dimensional momentum
cut-off hypersphere and then add terms which will be designed to
produce a translationally invariant result. In order for this
program to be successful, one needs sufficient constraints to fix
the free parameters in order to arrive at a uniquely defined,
translationally
invariant answer. The parameters are fixed by eliminating worse
than logarithmically divergent terms, 
%(i.e., more precisely in the
%limit $\Lambda \longrightarrow \infty$ a translationally invariant
%result is produced)
 and by requiring consistency with perturbation theory 
dimensional regularization in the weak-coupling limit. 
In Sec.~\ref{sec_fermion} we present the formalism for the fermion DSE
and briefly summarize and compare the numerical studies for the renormalized
nonperturbative fermion propagator using the two schemes.  
In Sec.~\ref{sec_general} we analyse translational invariance theoretically 
for perturbative massless and massive QED$_4$. 
We also provide a derivation of the modified cut-off regularization 
scheme which agrees with  the translationally invariant dimensional 
regularization scheme. 
 Finally, in Sec.~\ref{sec_conclusions} we summarize and
conclude.

\section{Fermion Dyson-Schwinger Equation in Quenched QED$_4$}
\label{sec_fermion}

The DSE for fermion propagator in quenched QED$_4$ can be
represented diagramatically as~:

\begin{figure}[h]
\begin{center}
\refstepcounter{figure}
\addtocounter{figure}{-1}
%\epsffile{fig2.ps}
~\epsfig{file=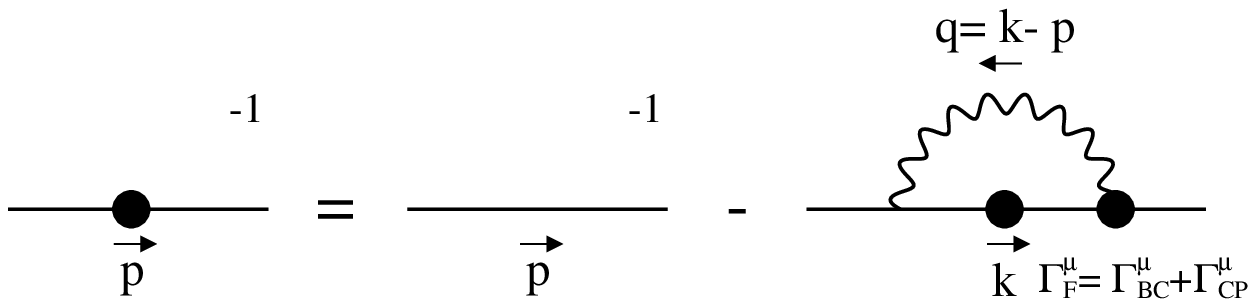,width=400pt}
\end{center}
%\hspace{4.2cm}
\caption{The inverse fermion propagator in quenched QED$_4$}
\label{fig:vertex}
%\vspace{-1cm}
%\vspace{-5cm}
\end{figure}

\noindent
Making use of the  Feynman rules for  this diagram leads to~:

\be
-{\it i}{\it S}_F^{-1}=-{\it i}{{\it S}_F^0}^{-1}
-\int_M\frac{d^4k}{(2\,\pi)^4}\, (-{\it i}e \Gamma^{\mu}_F)\, {\it
i}S_F(k^2)\, (-{\it i}e\gamma^{\nu})\,{\it
i}\Delta_{\mu\nu}(q^2)\quad, \label{eq:mainfer} \ee

\noindent
where introducing several frequently used notations at once~:

{\it{The Full Fermion Propagator}} is~:
\be
iS_F \equiv i\frac{F(p^2)}{\not\!p-M(p^2)}
     \equiv i\frac{Z(p^2)}{\not\!p-M(p^2)}
     \equiv i\frac{1}{A(p^2)\,\not\!p-B(p^2)} \quad,
\label{eq:fermionp}
\ee

\noindent
and $F(p^2)\equiv Z(p^2)\equiv 1/A(p^2)$ is the fermion wave-function
renormalization, $M(p^2) \equiv B(p^2)/A(p^2)\equiv Z(p^2)B(p^2)$
is the dynamical fermion mass.

\vspace{0.3cm}
{\it{ The Full Photon Propagator}} is~:

\be
i\Delta_{\mu\nu}(p^2)=-\frac{i}{p^2}\,
\left[G(p^2)\,
\left(g_{\mu\nu}-\frac{p_{\mu}p_{\nu}}{p^2}\right)
+\xi\frac{p_{\mu}p_{\nu}}{p^2}
\right]  \quad,
\label{eq:phoprop}
\ee

\noindent where $G(p^2)\equiv 1/(1+\Pi(p^2))$  is the boson
wave-function renormalization which is $1$ in the quenched
approximation, and $\xi$ the covariant gauge parameter. We shall
call
$\left(g_{\mu\nu}-p_{\mu}p_{\nu}/p^2\right) $
 the  {\it {transverse  part}} and
$p_{\mu}\, p_{\nu}/p^2$ the  {\it {longitudinal part}}.
Finally, $\Gamma^{\mu}$ is {\it {the full fermion-boson vertex}}
for which we use the CP
ansatz, namely ($q \equiv k - p$)
\be \label{anyfullG_eqn}
  \Gamma^\mu(k,p,q) = \Gamma_{\rm BC}^\mu(k,p)
    + \tau_{\rm CP}(k^2,p^2) \left [\gamma^{\mu}(p^2-k^2)+(p+k)^{\mu}
{\not \! q}\right ]\:,
\ee
where $\Gamma_{\rm BC}$ is the usual Ball-Chiu part of the vertex
which satisfies the
Ward-Takahashi identity~\cite{WTI}
\begin{eqnarray} \label{minBCvert_eqn}
  \Gamma^\mu_{\rm BC}(k,p) &=& \frac{1}{2}
\left(\frac{1}{F(k^2)}
+\frac{1}{F(p^2)}\right)
\gamma^\mu
   \\ \nonumber
& +& \frac{(k+p)^\mu}{k^2-p^2}
      \left\{ \left(\frac{1}{F(k^2)} - \frac{1}{F(p^2)}\right)
\frac{({\not\!k}+ {\not\!p})}{2}
              - \left(\frac{M(k^2)}{F(k^2)} -\frac{M(p^2)}{F(p^2)}\right)
\right\}\quad,
\ee
and the coefficient function $\tau_6$ is that chosen by Curtis and Pennington,
i.e.,
\be
  \tau_{\rm CP}(k^2,p^2) =
-\frac{1}{2\,d(k,p)}\left(\frac{1}{F(k^2)} -\frac{1}{F(p^2)}\right)\:,
\label{CPgamma1}
\ee
where
\be
  d(k,p) = \frac{(k^2 - p^2)^2 +
[M^2(k^2)  + M^2(p^2)  ]^2}{k^2+p^2}\:.
  \label{CPgamma2}
\ee

For studies of the nonperturbative renormalized fermion propagator,
dynamical chiral symmetry breaking, and  critical coupling
$\alpha_c$ in quenched QED$_4$, it is necessary to choose a specific
form for the fermion-photon proper vertex. While the exact form
of this vertex is not known, there are constraints and symmetries
which strongly restrict the allowable form.  These include
the Ward-Takahashi identity (WTI)\cite{WTI}), the absence of
artificial kinematic singularities, the requirements of
multiplicative
renormalizability (MR), and the need to agree with perturbation
theory in the weak coupling limit.  Furthermore, 
one should eventually ensure that
the gauge dependence of the resulting fermion propagator 
is consistent with the Landau-Khalatnikov transformation \cite{LKTF}
and that the value of the dynamical chiral symmetry breaking
critical coupling ($\alpha_c$) should be
gauge independent.

A number of discussions of the choice of the transverse part of the
proper vertex can be found in the literature, e.g.,
Refs.~\cite{BC,CPI,CPII,CPIII,CPIV,dongroberts,BP1,BP2,Kiz_et_al}.
We consider for our illustration here only the Curtis-Pennington
(CP) vertex\cite{CPI,CPII,CPIII,CPIV,ABGPR}, which
satisfies both the WTI and the constraints of multiplicative
renormalizability.  It is known that this vertex is
not entirely adequate to ensure the gauge invariance of $\alpha_c$,
although it is superior in this regard to the bare vertex
for example\cite{CPIV,ABGPR}.  The choice of a
vertex satisfying the WTI is a necessary but not sufficient condition
to ensure the full gauge covariance of the Green's functions of
the theory and the gauge invariance of physical observables.
Bashir and Pennington~\cite{BP1,BP2} have proposed alternatives
to the CP vertex, which ensure by construction that the critical
coupling indeed becomes strictly gauge independent.
All of the above studies were carried out with a UV cut-off
regularization, where it was necessary to remove by hand an
obvious explicit gauge-invariance violating term~\cite{dongroberts}
which arose
from the cut-off itself, (see for example
Ref.~\cite{qed4_hrw} for a detailed discussion of this).
We emphasize that the issue of interest here is not the construction
of the ideal
fermion-photon proper vertex, but rather to understand
and remove explicit gauge-invariance violations arising from
the UV cut-off regulator itself. For that purpose the CP vertex
provides a useful illustrative example.

%%%%%%%%%%%%%%%%%%%%%%%%%%%%%%%%%%%%%%%%%%%%%%%%%%%%%%%%%%%%%%%%%%%%%%%%%
%%%%%%%%%%%%%%%%%%%%%%%%%%%%%%%%%%%%%%%%%%%%%%%%%%%%%%%%%%%%%%%%%%%%%%%%
%%%%%%%%%%%%%%%%%%%%%%%%%%%%%%%%%%%%%%%%%%%%%%%%%%%%%%%%%%%%%%%%%%%%%%%

\subsection{Numerical Studies of the Fermion DSE}
\label{subsec_numerical}

In quenched QED there is no renormalization of the
electron charge and the appropriate photon propagator is just the bare
one.  The resulting nonlinear integral equation for the fermion
propagator is solved numerically.
The general features of dimensional regularization can be found
in any recent textbook (e.g., Ref.~\cite{Muta}), and we use the
notation $D = 4 - 2 \epsilon < 4$ for the dimension of Euclidean
space.  Successive calculations with decreasing $\epsilon$
must be numerically extrapolated to $\epsilon=0$.  For the UV cut-off
regularization we simply cut-off the Euclidean four-dimensional
loop integral at $\Lambda$ and verify that $\Lambda$ is sufficiently large.
 Clearly there is an ambiguity
about exactly which loop momentum variable has the cut-off applied
to it. In all calculations to date which use a UV cut-off a formal numerical
extrapolation to $\Lambda \longrightarrow \infty$ has not been
necessary.

The formalism is presented in
Minkowski space and the Wick rotation into Euclidean space can then
be performed once the equations to be solved have been written down.
Although we use dimensional regularization,
we cannot make use of the popular {\it perturbative} renormalization
schemes such as $MS$ or $\overline{MS}$, since they cannot be applied
in a nonperturbative context.

In the following equations and definitions, we will use $\epsilon$
to denote generic regularization dependence, where we must take the
generic $\epsilon\to 0$
in the limit where the regularization is removed.  For the UV
cut-off case we understand $\epsilon\sim 1/\Lambda$.
The renormalized inverse fermion propagator is defined through
\begin{eqnarray}                        \label{fermprop_formal}
            S^{-1}_R(\mu;p)  = \frac{1}{F_R(\mu;p^2)} \not\!p
                             - \frac{M_R(\mu;p^2)}{F_R(\mu;p^2)}
            & = & Z_2(\mu,\epsilon) [\not\!p - m_0(\epsilon)]
                              - \Sigma_0(\mu,\epsilon; p) \quad,\nonumber\\
            & = & \not\!p - m(\mu) - {\Sigma}_R(\mu;p)\;\;\;,
\end{eqnarray}
where $\mu$ is the chosen renormalization scale, $m(\mu)$ is the value
of the renormalized mass at $p^2 = \mu^2$, $m_0(\epsilon)$ is the bare
mass and $Z_2(\mu,\epsilon)$ is the
wave-function renormalization constant.
Due to the WTI for the fermion-photon
proper vertex, we have for the vertex renormalization constant
$Z_1(\mu,\epsilon)=Z_2(\mu,\epsilon)$. The renormalized and unrenormalized
fermion self-energies are denoted as ${\Sigma}_R(\mu;p)$
and $\Sigma_0(\mu,\epsilon;p)$ respectively.  These can be expressed in terms
of Dirac and scalar pieces
\begin{equation}
  \Sigma_0(\mu,\epsilon; p) = \Sigma^d_0(\mu,\epsilon; p^2) \not\!p
                     + \Sigma^s_0(\mu,\epsilon; p^2)\;\;\;,
  \label{decompose}
\end{equation}
and similarly for ${\Sigma}_R(\mu;p)$. We do not explicitly
indicate the dependence on $\epsilon$ of the renormalized
quantities $1/F_R(\mu;p^2)$, $M_R(\mu;p^2)/F_R(\mu;p^2)$ and
${\Sigma}_R(\mu;p)$, since for renormalized quantities we will
always be interested in their $\epsilon\to 0$ limit.  The
renormalized mass function $M(p^2)$ is renormalization point
independent due to the nature of multiplicative renormalizability
\cite{qed4_hrw}. The renormalization point boundary condition
\begin{equation}
  \left. S_R^{-1}(\mu;p) \right|_{p^2 = \mu^2}
  = \not\!p - m(\mu)\quad,
\label{ren_point_BC}
\end{equation}
implies that $F_R(\mu;\mu^2) \equiv 1$ and $m(\mu) \equiv M(\mu^2)$ and
gives
\begin{equation}\label{ren_BC}
  {\Sigma}^{d,s}_R(\mu; p^2) =
    \Sigma_0^{d,s}(\mu,\epsilon; p^2) - \Sigma_0^{d,s}(\mu,\epsilon; \mu^2)
     \quad.
\end{equation}
Also, the wave-function renormalization is given by
\begin{equation}
  Z_2(\mu,\epsilon) = 1 + \Sigma_0^d(\mu,\epsilon; \mu^2)\quad,
\label{eq_Z2}
\end{equation}
and for the bare mass $m_0(\epsilon)$
\begin{equation}
  m_0(\epsilon) = \left[ m(\mu) - \Sigma_0^s(\mu,\epsilon; \mu^2) \right]
        / Z_2(\mu,\epsilon)\;\;\;.
\label{baremass}
\end{equation}
Under a renormalization point transformation $\mu \to \mu^\prime$,
$m(\mu^\prime) = M({\mu^\prime}^2)$ and $Z_2(\mu^\prime,\epsilon)
= Z_2(\mu,\epsilon)/F(\mu^\prime; \mu^2)$ as discussed in
Ref.~\cite{qed4_hrw}. Since we are working here in the quenched
approximation we have $Z_3(\mu,\epsilon)=1$, $e_0\equiv e(\mu)$,
and the  photon propagator ${\Delta}^{\mu \nu}(\mu;q)$ has its
perturbative form where $G(p^2)=1$.

The unrenormalized self-energy is given by the integral
\begin{equation} \label{reg_Sigma}
  \Sigma_0(\mu,\epsilon; p) = i \, (e(\mu) \nu^\epsilon)^2
    \int \frac{d^Dk}{(2\pi)^D} \gamma^\lambda {S}(\mu;k)
      {\Gamma}^\nu(\mu; k,p)
      {\Delta}_{\lambda \nu}(\mu;p-k)\:,
\end{equation}
where $\nu$ is an arbitrary mass scale introduced in
$D\equiv 4-2\epsilon$
dimensions  so that the renormalized coupling $e(\mu)$
remains dimensionless in the dimensional regularization
scheme.  For the UV cut-off case we have no need of $\nu$
since $\epsilon=0$ but instead we integrate over a four-dimensional
sphere whose radius is the UV cut-off $\Lambda$. The center of this
sphere is often taken to be $k_\mu=0$, although one could equally
well choose it to be at any location, e.g., at any
$(k_\mu + c\,p_\mu + b_{\mu}) =0$, where $c$ is an arbitrary
real constant
and $b_{\mu}$ is an arbitrary Euclidean four-momentum.

\subsection{Numerical Comparison
%between Cutoff and Dimensional Regularization of Fermion DSE
}

In Ref.~\cite{Schreiber:1998},  the renormalized dimensionally
regularized fermion DSE for the Curtis-Pennington vertex in
quenched approximation was studied numerically. Therein, it was
noted that the fermion propagator extrapolated to $\epsilon = 0$
using dimensional regularization differed from that obtained from
using a cut-off regulator `as is', but agreed with that obtained
by using a cut-off regulator with the modification proposed by
Ref.~\cite{dongroberts}, within the numerical accuracy of the study.
This was
observed for a massive solution with the coupling $\alpha=1.5$ and
the gauge parameter $\xi=0.25$. Here we explore whether this
agreement holds at much increased numerical precision for the more 
numerically tractable case of
$\alpha=0.6$ in a variety of gauges for both massive and massless
solutions of the quenched fermion DSE.

\subsubsection{ Massive Case}

Fig~\ref{sub25eps.eps} shows a family of solutions calculated in
dimensional regularization scheme with the regulator parameter $\epsilon$ 
decreased from 0.08 to 0.03 for the coupling $\alpha=0.6$. The gauge
parameter is $\xi=0.25$, the renormalization point $\mu^2=10^8$ and the
renormalized mass is $m(\mu)=400$.  [Note that we have chosen our units such
that $\mu^2=10^8$ and $m(\mu)=400$ in those units.  We could equally well
choose these units to be MeV, eV, GeV, etc.  For a given solution
we can simply multiply all mass scales in the problem by the same
arbitrary constant and we still have a valid solution].
It is important to note the strong
dependence of the solutions on $\epsilon$, even though this parameter is
already rather small. The ultraviolet is most sensitive to this regulator,
however even in the infrared there is considerable dependence due to the
intrinsic coupling between these regions by the renormalization procedure.
This strong dependence on $\epsilon$ should be contrasted with the situation
in cut-off based studies where at rather modest cut-offs ($\Lambda^2 \approx
10^{10}$) the renormalized functions $A$ and $M$ had already reached 
 their asymptotic limits.

Also shown is the result of extrapolating these solutions to $\epsilon=0$ by
fitting a polynominal quartic in $\epsilon$ at each momentum point. As was
observed in \cite{Schreiber:1998}, the linearity and stability of the
extrapolation may be improved by a suitable choice of scale,$\nu$~: 
of the scales 1,
10, 100, 1000, and 10000, the latter two fit these criteria best. We show
$\nu=10000$ on the graphs.

Fig.~\ref{sub25fitins.eps} shows this extrapolated solution along with the
corresponding cut-off results, both with and without the aforementioned
modification which eliminates a spurious term induced by the cut-off which
breaks translational invariance.  As may be clearly seen from the insert of
Fig~\ref{sub25fitins.eps} the modified ultaviolet cut-off curve is
indisguishable from the  scaled dimensionally regularized one, while the
naive cut-off curve clearly deviates from the others in the infrared.
This observation is quantified by tables \ref{massive_cmp_a_table} and
\ref{massive_cmp_m_table} which show absolute percentage comparisons of the
finite renormalization and the mass function for the extrapolated solution
with the modified and naive UV cut-off massive solutions respectively, with
parameters as in Fig.~\ref{sub25eps.eps} for two different scales and two
different polynomial degrees. The agreement with the modified cut-off solution
is seen to be excellent, and is three orders of magnitude better than the
agreement with the naive cut-off.

Finally, Fig~\ref{submfitins.eps} shows a comparison in three different gauges,
namely $\xi=0$, $\xi=0.25$ and $\xi=1$, of solutions of the fermion
DSE extrapolated to $\epsilon=0$, and solutions using naive and modified UV
cut-off regulators, with other parameters the same as in 
Fig.~\ref{sub25eps.eps}. The $A(p^2)$ solutions are identical in
Landau gauge, and
the agreement between the extrapolated solution and the modified cut-off
solution is readily distinguished in Feynman gauge. Owing to the approximate
gauge invariance of the mass function, an insert is neccessary to reveal the
same holds for $M(p^2)$.

\subsubsection{Massless Case}

Fig~\ref{subz25epscloser.eps} shows a family of solutions calculated in
the dimensional regularization scheme with the regulator parameter $\epsilon$
from from 0.04 to 0.005 for the coupling $\alpha=0.6$, the gauge parameter
$\xi=0.25$ and the renormalization point $\mu^2=10^8$.  As in the massive
case, there is a strong dependence of the renormalized function $A(p^2)$ on the
regulator $\epsilon$, although here the infrared is even more sensitive than
the ultraviolet to $\epsilon$. This contrasts with cut-off solutions which
reach their asymptotic limit at rather modest cut-offs.

Also shown is the result of extrapolating these solutions to $\epsilon=0$ by
fitting a polynominal quartic in $\epsilon$ at each momentum point to
$\log_{10}(A)$. This was appropriate because the the
logarithmically scaled axes reveals the power-law character of the
extrapolated solution. 

In Fig~\ref{subz25fitins.eps} we show this extrapolated solution 
along with curves of the naive and modified cut-off solutions based
on the power-behaved analytical formulae 
Eqs.~(\ref{massless-naive-cutoff-soln})
and (\ref{massless-mod-cutoff-soln}) \cite{CPI,Brown:1991,BKP}
discussed in the next section of the form
\be
	F(p^2)=\left(p^2/\Lambda^2\right)^\gamma \, .
\ee
The curves are indistinguishable on the main figure: an insert reveals the
extrapolated solution agrees with the modified cut-off solution. This
is quantified by table \ref{massless_cmp_table} which shows absolute
percentage comparisons of $A(p^2)$ for these solutions. As in the massive
case, the agreement with the modified cut-off solution is many orders of
magnitude better than that with the naive cut-off.

Finally Fig~\ref{subzfit.eps} shows a comparison in three different
gauges, namely $\xi=0$, $\xi=0.25$ and $\xi=1$, of solutions of the fermion
DSE extrapolated to $\epsilon=0$, and solutions using naive and modified UV
cut-off regulators, with other parameters the same as in 
Fig.~\ref{subz25epscloser.eps}. 
As in the massive case, the $A(p^2)$ solutions are
identical in Landau gauge, and the agreement between the extrapolated and
modified cut-off solution is clear in Feynman gauge.

%%%%%%%%%%%%%%%%%%%%%%%%%%%%%%%%%%%%%%%%%%%%%%%%%%%%%%%%%%%%%%%%%%%%%%%%%%
%%%%%%%%%%%%%%%%%%%%%%%%%%%%%%%%%%%%%%%%%%%%%%%%%%%%%%%%%%%%%%%%%%%%%%%%%%%%

\section{Theoretical Studies on Translational Invariance of the DSE}
\label{sec_general}

The DSE approach to calculating any nonperturbative renormalized
Green's function in a renormalizable quantum field theory involves
an integral over a loop momentum, where the integrand involves two
or more renormalized nonperturbative Green
functions\cite{DSE_Review}.  If the Green's function to be
calculated from the loop integral corresponds to one of the
primitively divergent diagrams, then regularization of the loop
integration and renormalization will in general be necessary. The
primitively divergent diagrams in QED are the 2 and 3-point Green
functions, i.e., the fermion and photon propagators and the
fermion-photon proper (i.e., one-particle irreducible) vertex. For
other higher $n$-point Green's functions the loop integrations in
the DSE formalism are necessarily finite and renormalization of
these quantities is not needed once the nonperturbative
primitively divergent diagrams have been renormalized.

For instance the fermion self energy part of Eq.~(\ref{eq:mainfer}),
as it stands, has a linear ultraviolet
divergence because the integrand behaves like
$ \int \frac{d^4k}{k^3}$ for large $k$. Therefore, an
ultraviolet regulator must be introduced in order to perform the
loop integral.
%separate the divergent part (regulator dependent part) of the
%integral from the finite part(regulator independent part).
When the regulator is removed
by renormalizing the theory, one should be left with a
finite quantity which is
independent of the scheme used. If two regularization schemes give
different results then  some symmetries have been violated by one
(or both) of the regulators .
The regulatization scheme
should be chosen carefully so that  gauge invariance
and Poincare symmmetry in QED are preserved.
For instance, while the Pauli Villars  and dimensional
regularization schemes respect gauge and translational invariance, UV cut-off
regularization does not. However, one can attempt to use a UV cut-off
regulator
and still preserve these symmetries by imposing them on the regulator itself.
Whether or not this procedure will be unique and conserve
{\underline{\it all}} symmetries is the key question.
%demand these symmetries by imposing them on the regulator but
%of course, uniqueness and
% whether all symmetries are conserved is  subject to discussion
%or (can not be proven).
Here, a translationally invariant regularization scheme
is defined to mean that the same results are obtained
after arbitrary shifts in the definition of the loop
mometum variable in the limit that the regularization
is removed.  Since a UV cut-off regularization is a
restriction of the Euclidean loop-momentum integral
to a four-dimensional hypersphere of radius $\Lambda$,
we wish to ensure that in the limit $\Lambda\to\infty$
we find that the results are insensitive to the
location of the centre of this hypersphere.

We are interested in establishing  necessary and
sufficient conditions for a UV cut-off regularization scheme
to reproduce the results of a dimensional regularization
scheme for the renormalized $n$-point Green's functions. The scope of this
present work is to establish a procedure for this for the electron
self-energy.
The procedures for removing unwanted contributions in
a UV cut-off scheme are straightforward:
(1)
%Examine the loop momentum integrand and remove
%any terms that would vanish in a translationally
%invariant regularization scheme.
The best way to begin
identifying such terms is to  replace nonperturbative
quantities in the integrand by their perturbative form;
(2) Test that the resulting expression for the integral of
the nonperturbative renormalized quantity is independent of
the location of the hyperspheres center in the limit
$\Lambda\to\infty$.  Let us begin with an analysis of perturbation theory
 in order to understand the problem~:

\subsection{Perturbation Theory as a Guide}
\label{subsec_perturbative}
%
%
%
%\vspace{5cm}
\begin{figure}[h]
\begin{center}
\refstepcounter{figure}
\addtocounter{figure}{-1}
%\epsffile{fig2.ps}
~\epsfig{file=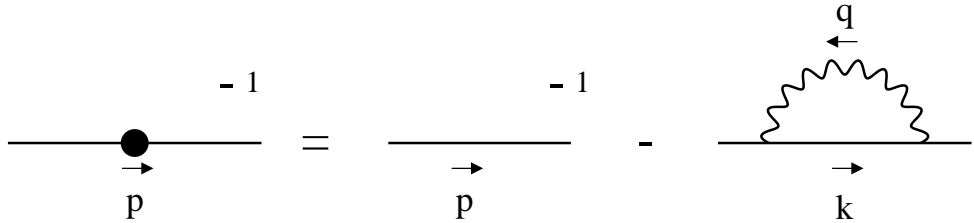,width=400pt}
\end{center}
%\hspace{4.2cm}
\caption{The inverse fermion propagator to one loop order in perturbation
theory}
\label{fig:vertwo}
%\vspace{-1cm}
%\vspace{-5cm}
\end{figure}
\noindent
Taking the weak coupling limit of Eq.~(\ref{eq:mainfer}) for massless
QED$_4$ up to
$ {\cal{O}}(\alpha)$, substituting in the fermion and
photon propagators and
the 3-point vertex
function, multiplying it by $\not\!p$ and taking its trace leads to  the
following expression~:

\small
\be
\frac{1}{F(p^2)}
=1+\frac{i\,\alpha}{4\,\pi^3p^2}\,\int_M
\frac{d^4k}{k^2q^4}\,
\Big\{
\left(
-2\,k\cdot p\,q^2
 \right)
+
(\xi-1)\,
\left(
k^2\,p\cdot q-p^2\,k\cdot q
 \right)
\Big\} \quad,\hspace{2.5cm}
% \\ [5mm]
\label{eq:perfermion1}
\ee
\sbe
\hspace{0.9cm}
&=&1+\frac{i\,\alpha}{4\,\pi^3p^2}\,\int_M
\frac{d^4k}{k^2q^4}\,
\Big\{
\underbrace{
%\left(
-3\,k\cdot p\,(k^2+p^2)+4(k \cdot p)^2+2k^2p^2
% \right)
}
+
\xi\,
\underbrace{
\left(
k^2\,p\cdot q-p^2\,k\cdot q
 \right)}
\Big\} \quad.\\
%\label{eq:perfermion2}
%
%
%
&&
 \hspace{6.8cm} I_T \hspace{2.5cm}
+
 \hspace{2cm} I_L \hspace{0.5cm} \nonumber
%\label{eq:perfermion}
\see

\normalsize
\be
\label{eq:perfermion2}
\ee
where ``M'' denotes Minkowski space.
\normalsize
\vskip 1cm
\noindent
{\bf{\underline{ ANALYSIS}}}:
%%%%%%%%%%%%%%%%%%%%%%%%%%%%%%%%%%%%%%%%%%%%%%%%%%%%%%%%%%%%%%%%%%%%%
%%%%%%%%%%%%%%%%%%%%%%%%%%%%%%%%%%%%%%%%%%%%%%%%%%%%%%%%%%%%%%%%%%%

\noindent
{\bf (1)~:}

\noindent
First we shall calculate the fermion wave-function renormalization within
{\it {\bf Dimensional Regularization}}, which respects the symmetries
of QED.
It is important to note that the
first term and the coefficient of $(\xi-1)$ in the curly bracket of
Eq.~(\ref{eq:perfermion1}) give the same answer implying that
the non-${\xi}$ part, $I_T$, of Eq.~(\ref{eq:perfermion2})  vanishes and the
${\xi}$ part, $I_L$,  yields the result as below~:

\be
\frac{1}{F(p^2)}=1-\frac{\alpha\,\xi}{4\pi}\,\left[
                  \frac{-1}{\epsilon}+\ln\left(\frac{ p^2}{\nu^2}\right)
                -1+\gamma-\ln(4\,\pi)
                                                   \right]\quad.
\label{eq:dimreg}
\ee
%
%
%
%As one can see first term and the coefficient of $(\xi-1)$ term
%in Eqn.~(\ref{eq:mainfermion}) yields the same answer. Therefore,
%after cancellation between non-$\xi$ terms, the fermion self
%energy part becomes  proportional to $\xi$.
%%%%%%%%%%%%%%%%%%%%%%%%%%%%%%%%%%%%%%%%%%%%%%%%%%%%%%%%%%%%%%%%%%%%%
%%%%%%%%%%%%%%%%%%%%%%%%%%%%%%%%%%%%%%%%%%%%%%%%%%%%%%%%%%%%%%%%%%%

\noindent
{\bf (2)~:}

\noindent
Repeating the same calculation as above except this time using
 {\it{\bf Cut-off Regularization}} we get with the hypersphere centre at
$k_{\mu}=0$~:

\be
\frac{1}{F(p^2)}=1-\frac{\alpha\xi}{4\,\pi}\,
\left(\ln\frac{p^2}{\Lambda^2}-
\mbox{\fbox{
$\displaystyle{\bf\frac{1}{2}}$
}}
\,\,\right)\quad.
\label{eq:cutoff}
\ee

\noindent
Once again the integral of $I_T$ from Eq.~(\ref{eq:perfermion2}) is zero.
%The above expression comes from the term proportional to $ {\xi}$
% in Eqn.~(\ref{eq:perfermion2}),
%and again  the term proportional to non-${\xi}$ part i.e. $I_T$ vanishes.

The $1/2$ term in Eq.~(\ref{eq:cutoff}) is a consequence of cut-off
regularization not preserving
translational invariance and gauge covariance;
in other words it is due to the non-conservation
of current. The Landau-Khalatnikov (LK) transformations determines the gauge
covariance of the theory in that if any Green's functions of the theory are known
in one gauge they are  also known for any other gauge via this
transformation. If in the Landau gauge, the fermion wave-function
renormalization is
$F(p^2,\Lambda^2)=\,A_0\,(p^2/\Lambda^2)^{\gamma_0}$
then this transforms to the covariant gauge as
$ F(p^2,\Lambda^2)=\,A\,(p^2/\Lambda^2)^{\gamma}$ where
$\gamma=\,\gamma_0+\alpha\,\xi/4\pi$ and $A,A_0$ are constants,\cite{BKP}.
 $\gamma_0$
is gauge independent term and in perturbation theory of ${\cal{O}}(\alpha^2)$,
\cite{BKP}.
Therefore,
if one performs the perturbative expansion of 
$F(p^2)=\,A\,(p^2/\Lambda^2)^{\gamma}$
for small $\alpha$, then the $1/2$ term in  Eq.~(\ref{eq:cutoff})
should be absent
in order to ensure that the solution of the DSE
for fermion wave-function renormalization is LK covariant.

 As we shall see below this term can be removed by making use of the WTI or
symmetry properties.

\begin{itemize}

%%%%%%%%%%%%%%%%%%%%%%%%%%%%%%%%%%%%%%%%%%%%%%%%%%%%%%%%%%%%%%%%%%%%%%%%
%%%%%%%%%%%%%%%%%%%%%%%%%%%%%%%%%%%%%%%%%%%%%%%%%%%%%%%%%%%%%%%%%%%%%%%%

%%%%%%%%%%%%%%%%%%%%%%%%%%%%%%%%%%%%%%%%%%%%%%%%%%%%%%%%%%%%%%%%%%%%%
%%%%%%%%%%%%%%%%%%%%%%%%%%%%%%%%%%%%%%%%%%%%%%%%%%%%%%%%%%%%%%%%%%%%%%

\vskip 0.5cm

\item

\noindent
The WTI follows from gauge invariance. Applying it
to the $\xi$-part of photon propagator in
Eq.~(\ref{eq:mainfer}),  separates
out the term, $p\cdot q/q^4$, which is zero in any translational invariant
scheme (odd in $q$).
On the other hand, in  cut-off regularization
it is the source of the  $1/2$ term in Eq.~(\ref{eq:cutoff}).

\vskip 0.5cm

%%%%%%%%%%%%%%%%%%%%%%%%%%%%%%%%%%%%%%%%%%%%%%%%%%%%%%%%%%%%%%%%%%%%%%%%
%%%%%%%%%%%%%%%%%%%%%%%%%%%%%%%%%%%%%%%%%%%%%%%%%%%%%%%%%%%%%%%%%%%%%%%%
\vskip 0.5cm

\item

\noindent
To analyse the  {\it{ translational invariance}}\, in cut-off regularization
we shall
shift the centre of the sphere
from $k_{\mu}=0$ to
$k_{\mu}=c\,p_{\mu}$, where $c$ is an arbitrary real constant,
in Eq.~(\ref{eq:perfermion1}) and Eq.~(\ref{eq:perfermion2}). In so
doing we obtain the following equations ~:

\small
\be
\int^{\Lambda} d^4k \,
\frac{-2\,k\cdot p}{k^2\, (k-p)^2}
&=&\pi^2p^2
\left\{\ln\frac{p^2}{\Lambda^2}-\frac{1}{2}\right\} \nonumber \\
&& \hspace{10 mm}
 \mathop{\longrightarrow}^{k \to k-cp}\qquad
%\overrightarrow{k \to k+np}
\pi^2p^2
\left\{\ln\frac{p^2}{\Lambda^2}+ \left(-\frac{1}{2}+c\right)
+{\cal{O}}(\Lambda^{-1})\right\} \quad,
\label{eq:tr1} \\
\int^{\Lambda} d^4k \, \frac{- k\cdot q\,p^2}{k^2\, (k-p)^4}
&=&\pi^2p^2\left\{ \ln\frac{p^2}{\Lambda^2}\right\} \nonumber \\
&& \hspace{10 mm}
\mathop{\longrightarrow}^{k \to k-cp}\qquad
\pi^2p^2\left\{ \ln\frac{p^2}{\Lambda^2}+{\cal{O}}(\Lambda^{-1})\right\}
\quad,
\label{eq:tr2}\\
\int^{\Lambda} d^4k \, \frac{p\cdot q}{ (k-p)^4}
&=&\pi^2p^2\left\{-\frac{1}{2}\right\} \nonumber \\
&& \label{eq:tr3} \hspace{10 mm}
\mathop{\longrightarrow}^{k \to k-cp}\qquad
\pi^2p^2\left\{-\frac{(c+1)}{2}\right\} \quad,
\ee
\be
\int^{\Lambda} d^4k \,
\frac{-3\,k\cdot p\,(k^2+p^2)+4(k \cdot p)^2+2k^2p^2}{k^2\, (k-p)^4}
&=& 0  \nonumber \\
%
%
%
%&& \hspace{10 mm}
\mathop{\longrightarrow}^{k \to k-cp}\qquad
&&
%\overrightarrow{k \to k+np}
\pi^2p^2
\left\{ \frac{3\,c}{2}
+{\cal{O}}(\Lambda^{-1})\right\} \quad.
\label{eq:tr4}
\ee
\normalsize

Examination of Eqs.~(\ref{eq:tr1})-~(\ref{eq:tr4}) reveals that with the
exception of Eq.~(\ref{eq:tr2}) (which is logarithmically divergent)
all others are  linearly divergent and VIOLATE  translational invariance~:

Looking at the first term in $I_L$, which corresponds to Eq.~(\ref{eq:tr3}),
 the WTI helps us to immediately recognize it as odd
and linearly divergent.
In cut-off regularization the position of the sphere is very important for
the regularized quantities.
If it is placed at $k_{\mu}=0$ then the $1/2$ term is generated in
Eq.~(\ref{eq:cutoff}) and Eq.~(\ref{eq:tr3}) which is consequence
of the violation of translational invariance. On the other hand if
the centre is located at $k_{\mu}=- p_{\mu}$ i.e. $c=-1$ then
Eq.~(\ref{eq:tr3})
is zero, translational invariance is preserved and cut-off is
consistent with dimensional regularization.
%both dimensional regularization and cut-off are consistent.
%
%As was disscussed earlier this term must be
%removed otherwise it will cause a  difference between shifted and
%unshifted integrals by an extra constant breaking translational invariance.
%In other words, this correcponds to $n$ being ${\bf{n=-1}}$ in
%Eqn.~(\ref{eq:tr3}).

The second term in $I_L$ has  only a logarithmic divergence and when it is
 shifted  arbitrarily, the difference between the shifted and unshifted
value  vanishes as $\Lambda \longrightarrow \infty$
so this term is translational invariant even under cut-off regularization,
Eq.~(\ref{eq:tr2}).

Finally we shall consider the transverse part, $I_T$ of
Eq.~(\ref{eq:perfermion2}) or Eq.~(\ref{eq:tr4}).
For large $k$ it becomes~:

\be
I_T&=& \int d^4k \left\{
-\frac{3\,k \cdot p}{k^4}+\frac{1}{k^6}\,
%&&
\underbrace{
\left(-4(k \cdot p)^2+k^2p^2\right)
}
\right\} \quad,
\label{eq:expan}\\
&& \hspace{5.6 cm} 0 \nonumber
\ee

as can be seen  this expression,
 is zero (convergent) in any translationally invariant scheme.
 The reason for this is that  the integral of the
linearly divergent (also odd in $k$) term ($-3\,k \cdot p/k^4$)
  is zero and the logarithmically divergent  second
and third terms  cancel each other out in any translationally invariant 
scheme, since they cancel after the angular integration about $k_\mu=0$.
If Eq.~(\ref{eq:expan}) is centred at $k_{\mu}=0$ then  even in cut-off
regularization, the integral of the first term is identically zero.
 Conversely, if the hypersphere is centred at $k_{\mu} \ne 0$ the
integral of the first term is non-zero. For instance, in
Eq.~(\ref{eq:tr4}), $k_{\mu}=c\,p_{\mu}$ will give $3\,c/2$.
Hence,
  in order to be consistent with the translational invariant regularization
$c$ must be ${\bf{zero}}$, i.e. when the transverse part is calculated one
should keep the centre at $k_{\mu}=0$ or one must add suitable terms to 
compansate.
Up until now,  the general
 covariant gauge case has been discussed.
However, often it is easier to calculate the fermion wave-function
renormalization  in a specific gauge. Take for example Feynman gauge,
$\xi=1$, in this case Eq.~(\ref{eq:perfermion1}) greatly simplifies
and we are left with only the first term of the integral,
Eq.~(\ref{eq:tr1}). From this equation we can see that in order
to be consistent with the arbitrary covariant gauge calculation and
dimensional regularization calculation $c$ can conveniently be chosen 
to be $1/2$.
In the arbitrary covariant gauge calculation, the
 non-$\xi$ part of the integral in
Eq.~(\ref{eq:perfermion1}) is given by
Eqs.~(\ref{eq:tr1}-\ref{eq:tr2}-\ref{eq:tr3}). In this case, the terms
violating translational invariance cancel out in Eq.~(\ref{eq:tr1}) and
Eq.~(\ref{eq:tr3}) if and only if $c=0$. In the Feynman gauge no
cancellation occurs so that the term violating the translational invariance
in Eq.~(\ref{eq:tr1}) must vanish identically. This can only be done
if the hypersphere is centred at $k_{\mu}=p_{\mu}/2$, namely $c=1/2$. We 
emphasize that we can equally well choose either centre for {\em any} 
gauge. The two choices described here give the same translationally
invariant result since the two choices can be seen to differ by just the
right term (that vanishes in a translationally invariant scheme).

To summarize:  In an arbitrary covariant gauge $I_T$ vanishes as it should,
if the centre of the hypersphere is located at $k_{\mu}=0$.
In that case, centering the cut-off integration for $I_L$ at
$k_{\mu}=q_{\mu}$ gives the result consistent with the translationally
invariant dimensional regularization.  This is one convenient
procedure.  In the specific case of the
Feynman gauge this same translationally invariant
result can also, for example, be conveniently  obtained by centering
 the hypersphere of the entire integrand at $k_{\mu}=p_{\mu}/2$.
These two ways of proceeding, of course, lead to the same expression
for the total integral in Feynman gauge as they should.
\end{itemize}

%%%%%%%%%%%%%%%%%%%%%%%%%%%%%%%%%%%%%%%%%%%%%%%%%%%%%%%%%%%%%%%%%%%%%
%%%%%%%%%%%%%%%%%%%%%%%%%%%%%%%%%%%%%%%%%%%%%%%%%%%%%%%%%%%%%%%%%%%%%%

\vspace{0.5cm}
\noindent
{\bf (3)~}
{\it {\bf Renormalization~:}}

{\bf{(a)~:}}
Applying
multiplicative renormalization (MR) requires the following relations
between renormalized, $F_R$, and unrenormalized, $F_0$, fermion wave-function
renormalization~:

\be
F_R=Z_2^{-1}\,F_0 \quad,
\label{eq:cutren}
\ee
where $Z_2$ denotes the fermion renormalization constant. If we apply
MR to the
perturbation theory order by order, then
the unrenormalized fermion wave-function renormalization which is calculated
in an uncorrected cut-off scheme  up to order ${\alpha}$ is~:

\be
F_0(p^2,\Lambda^2)&=&1+\frac{\alpha\,\xi}{4\,\pi}
\left(\ln\frac{p^2}{\Lambda^2}-\frac{1}{2}\right)+{\cal{O(\alpha^2)}}\quad,
%\nonumber \\
%
%
%
%&=&\left(1-\frac{\alpha}{2}+{\cal{O(\alpha)}}\,\right)
%\left(\frac{p^2}{\Lambda^2}\right)^{\frac{\alpha\,\xi}{4\,\pi}}
\ee

\noindent
and the fermion renormalization constant is~:

\be
Z_2(\mu^2,\Lambda^2)=1+\frac{\alpha\,\xi}{4\,\pi}
\left(\ln\frac{\mu^2}{\Lambda^2}-\frac{1}{2}\right)+{\cal{O}}(\alpha^2)
\ee
One can renormalize $F_0$ by choosing
$F_R(p^2,\mu^2)=1$ at  ${p^2=\mu^2}$
and find the renormalized
wave-function renormalization from Eq.~(\ref{eq:cutren}) as~:
\be
F_R(p^2,\mu^2)=1+\frac{\alpha\,\xi}{4\,\pi}
\ln\frac{p^2}{\mu^2}+{\cal{O}}(\alpha^2) \quad,
\ee
\noindent
which can be summed to all orders as~:
\be
F_R(p^2)=
\left(\frac{p^2}{\mu^2}\right)^{\frac{\alpha\,\xi}{4\,\pi}} \quad.
\ee

{\bf{(b)~:}}
Applying subtractive renormalization to the dimensional regularization
calculation, Eq.~(\ref{eq:dimreg}), we get~:
\be
{F_R(p^2)}&=&1+\Sigma_R(p^2;\mu^2) \\ \nonumber
&=& 1+\left(\Sigma_0(p^2;\mu^2,\Lambda^2)
-\Sigma_0(\mu^2;\mu^2,\Lambda^2)\right)
\\ \nonumber
&=&1+\frac{\alpha\xi}{4\,\pi}\ln\frac{p^2}{\mu^2}+{\cal{O}}(\alpha^2)\\
\nonumber
&=&\left(\frac{p^2}{\mu^2}\right)^{\frac{\alpha\xi}{4\pi}} \quad,
\ee
where $\Sigma_R(p^2;\mu^2)$ is renormalized fermion self energy.
As we see, even if we start with the incorrect unrenormalized fermion
wave-function renormalization, we get the same renormalized results for both
cut-off and dimensional regularization schemes in perturbation theory.
This is because the WTI
is taken care of automatically in perturbation theory;
however this is not the case
% but is not
in  nonperturbative theory. So, starting
with the wrong quantity, gives the wrong answer in nonperturbative theory.
%As we shall see, the
%difference in the results
If  one does not
impose translational invariance on the regulator as a necessary condition
then the result will be different in cut-off and dimensional
 regularization schemes.

Since the violation of translational
invariance in a naive UV cut-off scheme appears to be the only source
of gauge-covariance violation, the restoration of
translational invariance should also remove
any explicit source of gauge-covariance violation. 
 Translational invariance is certainly a necessary
condition, but one should ask whether it is a sufficient
condition.  At present we know of no rigorous mathematical
argument that proves such a sufficiency.  All that can be
said is that it is difficult to conceive of integrands
where this would not be the case.  Indeed,
field theories which yield different behaviors
from dimensional regularization and translationally
invariant UV cut-off approaches would need to be specified
by both a Lagrangian density {\em and} by a particular
choice of regularization scheme.

We now move on to nonperturbative QED and investigate how much information
 from perturbation theory we can make use of. Let us start with
Ball-Chiu (BC) plus Curtis-Pennington (CP) vertex in
massless QED$_4$.

%%%%%%%%%%%%%%%%%%%%%%%%%%%%%%%%%%%%%%%%%%%%%%%%%%%%%%%%%%%%%%%%%%%%%%%%
%%%%%%%%%%%%%%%%%%%%%%%%%%%%%%%%%%%%%%%%%%%%%%%%%%%%%%%%%%%%%%%%%%%%%%%%

%\subsection{Recipe for getting Translationally Invariant Result}
\subsection{Prescription for Consistency with Dimensional 
            Regularization}
 
Thus far we have seen how translational invariance is violated in the 
UV cut-off regularization and have described what must be done to obtain
the same result as dimensional regularization for the fermion
self-energy in quenched QED$_4$.
In this section we shall formulate a more general prescription.

In any translationally invariant regularization scheme the following
expression is true:

\be
&
{\footnotesize{{\mbox{translationally}}}}&
 \nonumber \\[-3mm]
&
{\footnotesize{\mbox{ invariant \, scheme }}} &\nonumber \\
I(p^2)=\int d^4k\, I(k,p,\theta)&=&
\int d^4k\, I(k',p,\theta)  \quad,\nonumber \\
\ee
\noindent
where the integrand $I(k,p,\theta)$ is related to the
shifted integrand
$I(k',p,\theta)$ through $k'_{\mu}=k_{\mu}+c\,p_{\mu}+b_{\mu}$.
 Unfortunately,  in cut-off regularization the above equality is not true
for renormalized integrands in the limit $\Lambda \longrightarrow \infty$
unless the
integrand, $I(k,p,\theta)$, is at worst logarithmically divergent,
i.e., in general

\be
{\mathop{\lim}_{ \Lambda \rightarrow \infty}}
\left(\int^{\Lambda} d^4k\, I(k,p,\theta)
-
\int^{\Lambda} d^4k\, I(k',p,\theta)\right)
\ne 0 \quad.
\label{eq:recipe}
\ee
Since in the limit $\Lambda\to\infty$ the renormalized Green
functions (and their various component parts) are necessarily
finite, the contribution from the integrands must be
vanishingly small at infinity.  This ensures that the result of
the integral is independent of whether the shape of the integral
region was hyperspherical, hypercubic or any other shape one might
construct. In other words, once translational invariance in the UV
cut-off approach has been ensured and the limit $\Lambda\to\infty$
has been taken, the resulting renormalized nonperturbative
quantities are entirely independent of the details of how the
limit was taken.
To ensure the above equality we need to develop an appropriate, unique, 
translationally invariant cut-off regularization scheme. This  can be
summarized by the following prescription~:

\vspace{0.3cm}
\renewcommand{\theenumi}{\Roman{enumi}}
\begin{enumerate}
%\renewcommand{\descriptionlabel}[1]%
%  {\hspace{\labelsep}\textsf{#1}}

%\begin{description}

\item
Start with the integral $\int^{\Lambda} d^4k \, I(k,p,\theta)$.

\item
Choose any centre for the four-dimensional hypersphere with radius $\Lambda$.

\item
Add  $\int d^4k\,\Delta I_i$ terms which would vanish in
any translational invariant
regularization scheme to eliminate the defect of all linearly divergent
terms in perturbation theory, where

\be
\int d^4k\,\Delta I_i
=\int d^4k\, \left[I_i(k',p,\theta)-I_i(k,p,\theta)\right] =\,0 \quad.
\ee
and where $I_i(k,p,\theta)$ is some integrand.
The above difference is zero in dimensional regularization but will
introduce an artificial term which contributes to the  next to leading
order terms in a cut-off scheme.
The number of  $\Delta I_i$ terms needed is related to the number of
linearly divergent terms in the integrand~,

\be
\int^{\Lambda} d^4k\,I'(k,p,\theta)=
\int^{\Lambda} d^4k\,\left[
 I(k,p,\theta)
+
 \sum_{i}\,d_i\,\Delta I_i(k,k',p,\theta)\right] \quad.
\label{eq:res3}
\ee
where $d_i$ are  constants.

\item
The $d_i$'s are to be fixed by equating Eq.~(\ref{eq:res3}) with the
translational invariant results (dimensional regularization results) 
from perturbation theory.
\end{enumerate}
\subsection{Application of the Prescription}

Let us  consider the fermion wave-function renormalization, $F(p^2)$, in
 perturbation theory as an example. As we shall disscuss later
one can see the residue of the violated symmetries in the cut-off scheme in
the nonperturbative case even after renormalization. Of course this is 
not the case
in perturbation theory. Therefore, since in this section we deal with
the perturbative
expansion of  the fermion wave-function renormalization, we shall use
regularized quantities in order to pin down the terms which cause problems
in the nonperturbative case.

The integrand in
Eq.~(\ref{eq:perfermion2}) can be divided into three parts~:

\be
I(k,p,\theta)=I_0(k,p,\theta)
+\,\xi\left[I{^{\rm odd}}(k,p,\theta)+\,I{^{\rm trans.}}(k,p,\theta)
\right]\quad,
\label{eq:applic}
\ee
where
\be
\left.
\begin{array}{cll}
\displaystyle I\,_0(k,p,\theta)&\equiv &\displaystyle \frac{\left(
-3\,k\cdot p\,(k^2+p^2)+4(k \cdot p)^2+2k^2p^2
 \right)}{k^2\,q^4}\quad, \nonumber \\
\displaystyle
I\,{^{\rm odd}}(k,p,\theta)&\equiv &
\displaystyle
\frac{k^2\,p\cdot q}{k^2\,q^4}\quad,  \\
\displaystyle
I\,{^{\rm trans.}}(k,p,\theta)&\equiv &
\displaystyle
\frac{-p^2\,k\cdot q}{k^2\,q^4}\quad,
\end{array}
\,\,\,\right\}
\ee
\noindent
where $I^{{\rm odd}}$ is odd in $q$ and $ I^{{\rm trans}}$ is a 
translationally
invariant integrand (since it is only logarithmically divergent).

\noindent
Let us apply the prescription
% Eqn.~(\ref{eq:shift})
to the ${\xi}$ part in Eq.~(\ref{eq:applic}) first~:

\renewcommand{\theenumi}{\Roman{enumi}}
\begin{enumerate}

\item
Start with,

\be
I_{\xi}(p) \equiv \int^{\Lambda} d^4k \left[
I\,{^{\rm odd}}(k,p,\theta)+\,I\,{^{\rm trans.}}(k,p,\theta)
\right] \quad.
\ee

\item
Choose $k_{\mu}$ as the centre of the hypersphere $\Lambda$.
\item
Add $\int d^4k\,\Delta I_i$ terms~:

\be
I'\,_{\xi}(p)&=&\int^{\Lambda} d^4k \left[
I'\,{^{\rm odd}}(k,p,\theta)+\,I'\,{^{\rm trans.}}(k,p,\theta)
\right]\quad,  \nonumber \\
&=&\int^{\Lambda} d^4k\, \Big\{\,\,
\,I{^{\rm odd}}(k,p,\theta)
+d_1\,\left[I{^{\rm odd}}(k',p,\theta)-I{^{\rm odd}}(k,p,\theta)\right]
\nonumber\\
&&+ \hspace{1.3cm}
I{^{\rm trans}}(k,p,\theta)
+d_2\,\left[I{^{\rm trans}}(k',p,\theta)-I{^{\rm trans}}(k,p,\theta)\right]
\Big\}\quad,
\label{eq:ksi1}
\ee
where $k'$ is the shifted $k$, in general
$k'_{\mu}=k_{\mu}+c p_{\mu}+b_{\mu}$.

\item
Fix $d_1$ by using the following~:
\end{enumerate}
\begin{itemize}

\item

{\it {In any translational invariant scheme, any odd part of the
integral should be zero}}~:

\be
\mbox{\fbox{$\displaystyle
\int d^4k I'\,{^{\rm odd}}(k,p)=\int d^4k I'\,{^{\rm odd}}(k',p)=0$}}
\ee

\noindent
Therefore the odd part of Eq.~(\ref{eq:ksi1}) can be written as~:

\be
I^{'\rm odd}(p)\,=\,0=\,\int^{\Lambda}d^4k\,\left\{
(1-d_1)\,I{^{\rm odd}}(k,p,\theta)+d_1\,I{^{\rm odd}}(k',p,\theta)
\right\}\quad.
\label{eq:as1}
\ee

\noindent
For instance, if we locate the centre of the sphere at $k_{\mu}=0$ and
then we shift the loop  momentum to
$k_{\mu}\longrightarrow k_{\mu}+p_{\mu}$ then~:

\be
\int^{\Lambda}d^4k\,I{^{\rm odd}}(k,p,\theta)&=&-\frac{1}{2}\quad,
 \nonumber \\
%
%
%\int^{\Lambda}d^4k\,I{^{\rm odd}}(k+p,p,\theta)=0\nonumber \\
%
%
\int^{\Lambda}d^4k\,I{^{\rm odd}}(k+p,p,\theta)&=&
\int^{\Lambda} d^4k\, \frac{(p \cdot k)}{k^4}=0 \quad.
\ee
%
%
%
%\Longrightarrow \quad m_1=1
%\ee
%
%
Hence, in order to satisfy Eq.~(\ref{eq:as1}),
%$ m_1$ should be~:
\sbe
d_1=1
\see

\item

{\it {The $I^{\rm trans}$ part is independent of shifts in $k$, i.e.
it is translationally invariant since it is only logarithmically divergent~:}}

\be
\mbox{\fbox{$\displaystyle
I^{'\rm trans}(p)=\int^{\Lambda}\,I{^{trans}}(k,p,\theta)\quad
=
\int^{\Lambda}\,I{^{trans}}(k',p,\theta) \quad\mbox{independent of $d_2$}
$}}
\ee

\noindent
Now let us consider the $I_0$ part of Eq.~(\ref{eq:applic})~:

\item

{\it {We know from dimensional regularization that in  perturbation theory}}
{\fbox{$\int d^4k\,I_0(k,p,\theta)=0$}}\,\,.

 \be
I_0'(p)=\,0=\,\int^{\Lambda}d^4k \left\{
(1-d_3)\,I_0(k,p,\theta)+d_3\,I_0(k',p,\theta)\right\}\quad.
\label{eq:as3}
\ee

Within cut-off regularization~:

\be
\int^{\Lambda} d^4k\,I_0(k,p,\theta)&=& 0 \quad,\\ \nonumber
\int^{\Lambda} d^4k\,I_0(k',p,\theta)&=&-\frac{3}{2}\quad,
\ee
where again here the first integral is centred at $k_{\mu}=0$ and the
second integral at $k_{\mu}=p_{\mu}$.
So, for  Eq.~(\ref{eq:as3}) to be true
%$m_3$ must be~:

\sbe
d_3&=&0 \quad.
\see

\end{itemize}

The above prescription for the cut-off regularization scheme should ensure
the same result
as the  dimensional regularization scheme.

\vspace{3cm}
%%%%%%%%%%%%%%%%%%%%%%%%%%%%%%%%%%%%%%%%%%%%%%%%%%%%%%%%%%%%%%%%%%%%%%%%%%%%%%
%%%%%%%%%%%%%%%%%%%%%%%%%%%%%%%%%%%%%%%%%%%%%%%%%%%%%%%%%%%%%%%%%%%%%%%%%%%%%
%%%%%%%%%%%%%%%%%%%%%%%%%%%%%%%%%%%%%%%%%%%%%%%%%%%%%%%%%%%%%%%%%%%%%%%%%%%%
\subsection{Massless, Quenched QED$_4$ with CP-Vertex }
\begin{figure}[h]
\begin{center}
\refstepcounter{figure}
\addtocounter{figure}{-1}
~\epsffile{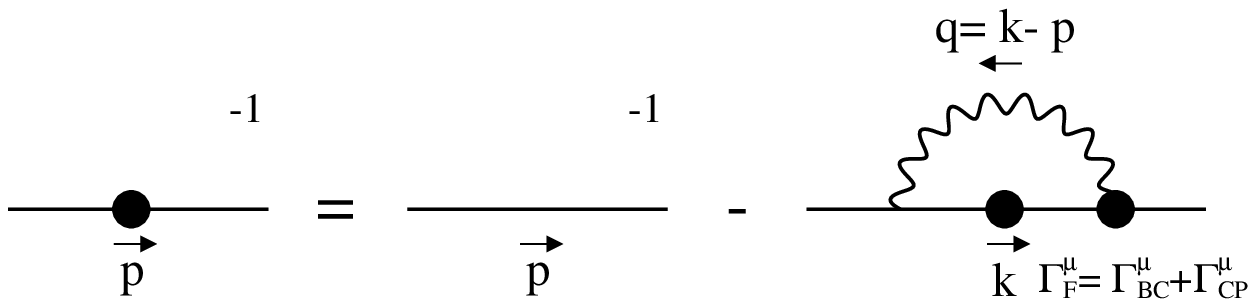}
\end{center}
\caption{Quenched Dyson-Schwinger Equation with Curtis-Pennington
Vertex} \label{fig:sdcp}
\end{figure}
%
%\vspace{-2cm}
%
Inserting the full fermion and bare photon propagators and  full
fermion-photon vertex (BC+CP) into  Eq.~(\ref{eq:mainfer}), then
multiplying the result by $\not\!p$, and taking its trace  we get~:

\footnotesize

\sbe
\frac{1}{F(p^2)}&=&1+\frac{i\,\alpha}{8\,\pi^3p^2}\,\int_M
\frac{d^4k}{k^2q^4}\,\times
\nonumber \\
&&\Bigg\{
\left(1+\frac{F(k^2)}{F(p^2)}\right)\,
\Bigg[\hspace{18mm}
\left(-3\,k\cdot p\,(k^2+p^2)+4(k \cdot p)^2+2k^2p^2\right)
+\xi\,\left(
k^2\,p \cdot q-p^2\,k\cdot q\right)
\Bigg] \nonumber \\
&&+
\left(1-\frac{F(k^2)}{F(p^2)}\right)\,
\Bigg[\frac{(k^2+p^2)}{(k^2-p^2)}\,
\left(-3\,k\cdot p\,(k^2+p^2)+4(k \cdot p)^2+2k^2p^2\right)
+\xi\,\left(
k^2\,p \cdot q+p^2\,k\cdot q\right)
\Bigg]
\Bigg\}\,.
\see
\be
\label{eq:bccp}
\ee

\normalsize

After moving from Minkowski space to Euclidean space by performing
a Wick rotation, we can carry out the above integrals.
By looking at Eq.~(\ref{eq:bccp}), we notice that the fermion wave-function,
$F(p^2)$,  is the
only nonperturbative quantity and does not depend on the angle between 
$k$ and $p$. At
the level of calculating angular integrals, everything is the same as for the
perturbation theory case, hence we know how to deal with these integrals.
As we have seen before, the first term of the integral in
Eq.~(\ref{eq:bccp}),
earlier  called $I_T$, is zero,( Eq.~(\ref{eq:tr4})). So then we have:

\be
\frac{1}{F(p^2,\Lambda^2)}=1-\frac{\alpha\,\xi}{4\,\pi^3}\,
\int_E\,\frac{d^4k}{k^2\,q^4}\,
\left(
k^2\,p\cdot q- p^2 k \cdot q\,\frac{F(k^2,\Lambda^2)}{F(p^2,\Lambda^2)}
\right) \, .
\label{eq:unfermion}
\ee

\noindent
The above equation can be expressed in terms of renormalized quantities as~:

\be
\frac{1}{Z_2(\mu^2,\Lambda^2)\,F_R(p^2,\mu^2)}=
1-\frac{\alpha\,\xi}{4\,\pi^3\,p^2}\,
\int_E\,\frac{d^4k}{k^2\,q^4}\,
\left(
k^2\,p\cdot q- p^2 k \cdot q\,\frac{F_R(k^2,\mu^2)}{F_R(p^2,\mu^2)}
\right) \quad.
\ee

\noindent
Multiplying this equation by $F_R(p^2,\mu^2)$ to leave
the fermion renormalization constant, $Z_2$, alone on the
left hand side
and performing some of the integrals, we find~:

\small
\sbe
\frac{1}{Z_2(\mu^2,\Lambda^2)}=
F_R(p^2,\mu^2)+\frac{\alpha\,\xi}{4\,\pi}\,
\int_0^{\Lambda^2}\,dk^2\,
\left(
 \frac{k^2}{p^4}\, F_R(p^2,\mu^2)\,\theta(p^2-k^2)
+ \frac{1}{k^2}\,F_R(k^2,\mu^2)\,\theta(k^2-p^2)
\right)\quad.
\see

\vspace{-0.5cm}

\be
\label{eq:peqmu}
\ee
\normalsize

\noindent
If we write the same expression for $p^2=\mu^2$ and subtract it from
Eq.~(\ref{eq:peqmu}), we get~:

\sbe
0=F_R(p^2,\mu^2)-F_R(\mu^2,\mu^2)
+\,\frac{\alpha\,\xi}{4\,\pi}\,
\Bigg\{\frac{1}{2}\,F_R(p^2,\mu^2)\,-\,\frac{1}{2}\,F_R(\mu^2,\mu^2)
+\int_{p^2}^{\mu^2}\,\frac{dk^2}{k^2}\,F_R(k^2,\mu^2)
\Bigg\}\quad.
\see

\vspace{-0.5cm}

\be
\label{eq:differ}
\ee

\noindent
Considering that the fermion wave-function renormalization $F(p^2)$
must obey the power behaviour $F_R(p^2,\mu^2)=(p^2/\mu^2)^{\nu}$
in the nonperturbative massless case, then after carrying out the
radial integral
Eq.~(\ref{eq:differ}) can be written as:

\be
F_R(p^2,\mu^2)-F_R(\mu^2,\mu^2)=-\frac{\alpha\,\xi}{4\,\pi}\,
\Bigg\{
\left(\frac{1}{2}-\frac{1}{\nu}\right)\,
\left[F_R(p^2,\mu^2)-\,F_R(\mu^2,\mu^2)\right]
\Bigg\}\quad.
\label{eq:cpsol}
\ee

\noindent
Comparing both sides of Eq.~(\ref{eq:cpsol}) we find
that\cite{ABGPR},\cite{BKP}
\be
\nu=\frac{2\alpha\xi}{\alpha\xi+8\pi}\quad \label{eq:nu} \ee
giving the solution
\be
F_R(p^2/\mu^2) &=&
\left(\frac{p^2}{\mu^2}\right)^{2\alpha\xi/(\alpha\xi+8\pi)}.
\label{massless-naive-cutoff-soln}
\ee

\noindent
 Unlike  perturbation theory, the third and fourth terms
of Eq.~(\ref{eq:differ}) do not cancel each other, so that the
first term, $1/2$, on the right hand side of Eq.~(\ref{eq:cpsol})
survives. We have seen in perturbation theory that keeping $\int
d^4k\, p\cdot q/q^4$
 which is the source of this term  did not make
any difference in the renormalized quantities,
because in perturbation theory it is~:

\be
\int d^4k\,\frac{p\cdot q}{q^4}
-\int d^4k\,\frac{\mu \cdot q'}{q'^4}=\,0 \quad,\\ \nonumber
\ee
where  $ q'=k-\mu$.
BUT it does make a difference in nonperturbative studies.
Assuming the transverse vertex vanishes  in the Landau gauge
means that $\nu=\alpha\,\xi/4\,\pi$, i.e.
$F_R(p^2,\mu^2)=(p^2/\mu^2)^{{\alpha\,\xi}/{4\,\pi}}$,
from the LKF transformation.
 Eq.~(\ref{eq:nu}) is
different from $\alpha\,\xi/4\,\pi$ due to the fact that translational
invariance is broken by cut-off regularization.
Therefore, one must cancel the  $\int d^4k\, p\cdot q/q^4$ term  in
Eq.~(\ref{eq:unfermion})
in order to recover the correct
behaviour of the fermion wave-function renormalization. Removing this
term from Eq.~(\ref{eq:unfermion}), we find~:
\be
\frac{1}{F(p^2,\Lambda^2)}=1+\frac{\alpha\,\xi}{4\,\pi^3}\,
\int_E\,\frac{d^4k}{k^2\,q^4}\,
\left(
 p^2 k \cdot q\,\frac{F(k^2,\Lambda^2)}{F(p^2,\Lambda^2)}
\right) \quad,
\label{eq:fermion}
\ee
so that
\be
\nu&=&\frac{\alpha\,\xi}{4\,\pi} \quad
\ee
giving the solution of this equation as
\be
F_R(p^2/\mu^2)&=&\left(\frac{p^2}{\mu^2}\right)^{\alpha\,\xi/4\,\pi}
\label{massless-mod-cutoff-soln}.
\ee

This is exactly the same as the result  obtained from a nonperturbative 
dimensional
regularization scheme \cite {dongroberts}. 
As a result of the above discussions we see that
the modified cut-off prescription can be succesfully applied to massless
quenched QED. Therefore, after applying the prescription for this case
one can see the agreement between  dimensional regularization and
the modified cut-off result numerically in Figs.\ref{subz25fitins.eps}
and \ref{subzfit.eps}.

%%%%%%%%%%%%%%%%%%%%%%%%%%%%%%%%%%%%%%%%%%%%%%%%%%%%%%%%%%%%%%%%%%%%%
%%%%%%%%%%%%%%%%%%%%%%%%%%%%%%%%%%%%%%%%%%%%%%%%%%%%%%%%%%%%%%%%%%%%%
\normalsize
\subsection{Massive, Quenched QED$_4$ with CP-Vertex }
The fermion wave-function renormalization
for the  massive QED$_4$ case  using BC and CP vertices can be written as~:

\small

\begin{eqnarray}
\frac{1}{F(p^2)}&=&1+\frac{i\,\alpha}{8\,\pi^3p^2}\,\int_M
\frac{d^4k}{\left(k^2-M^2(k^2)\right)\,q^4}\,\times \\
\nonumber
&&\Bigg\{
\left(1+\frac{F(k^2)}{F(p^2)}\right)\,
\Bigg[\hspace{18mm}
\left(-3\,k\cdot p\,\left(k^2+p^2\right)+4(k \cdot p)^2+2k^2p^2\right)
+\xi\,\left(
k^2\,p \cdot q-p^2\,k\cdot q\right)
\Bigg] \\ \nonumber
&&+
\left(1-\frac{F(k^2)}{F(p^2)}\right)\,
\Bigg[\frac{1}{\left(k^2-p^2\right)}\,\,\,
\frac{\left(-3\,k\cdot p\,\left(k^2+p^2\right)
+4(k \cdot p)^2+2k^2p^2\right)}{d}
+\xi\,\left(
k^2\,p \cdot q+p^2\,k\cdot q\right)
\Bigg] \nonumber \\
&&-
\left(1-\frac{F(k^2)}{F(p^2)}\right)\,
\Bigg[\frac{1}{\left(k^2-p^2\right)}\,
\frac{2\,\Delta^2\,\left(M^2(k^2)+M^2(p^2)\right)^2}{d}
\Bigg] \nonumber \\
%
%
%
%&&-
%\frac{2\,\Delta^2}{\left(k^2-p^2\right)}\,
%\Bigg(
%M^2(k^2)-M(k^2)M(p^2)\frac{F(k^2)}{F(p^2)}
%+\frac{\left(M^2(k^2)+M^2(p^2)\right)^2}{2\,d_E}
%\Bigg) \nonumber \\
%
%
%
&&+
\left(M^2(k^2)-M(k^2)M(p^2)\frac{F(k^2)}{F(p^2)}\right)\,
\frac{1}{(k^2-p^2)}\,
\Bigg(
-4\Delta^2
+\xi\,(k^2-p^2)^2 p \cdot q
\Bigg)
\Bigg\} \quad,
\label{eq:massive}
\end{eqnarray}
\normalsize
where
\sbe
d&=&\frac{(k^2 - p^2)^2 +
[M^2(k^2)  + M^2(p^2)  ]^2}{k^2+p^2}\quad, \\
\Delta^2&=& (k \cdot p)^2-k^2p^2\quad.
\see
In this case, the second and third lines of Eq.~(\ref{eq:massive}) are
exactly the same as in the massless case except for the mass term in
the denominator.
Hence, for calculating these lines,
the only difference between the massless and massive
cases will come from radial integrals and
the presence of a mass term in
the denominator only makes the calculation convergent more quickly.
So we do not
encounter any worse  than a logarithmic divergence. The
third and fourth lines of Eq.~(\ref{eq:massive}) will introduce new terms
which depend on the mass function but the integrals do not have any
worse
divergence than the logarithmic divergence because for large momenta
the mass function  behaves like
$ M(k^2)\, \alpha \, k^{-\gamma}, 0<\gamma<2 $, \cite{ABGPR},
 \cite{qed4_hw},\cite{qed4_hrw}.
  Therefore, there is no danger of violating
translational invariance.

Of course, in the massive case with the Curtis-Pennington or
the real transverse
vertex, the fermion wave-function renormalization will not give $1$ in
Landau gauge. In other words, the transversality condition~\cite{trans},
\cite{Kiz_et_al}
is not applicable
 for these vertices. As a result of that, for such vertices, we can not use
the condition
$F(p^2)=1$ in Landau gauge in the prescription. Consequently, the third item,
Eq.~(\ref{eq:as3}),
in the prescription must be changed to~:

\be
I_0'(p)={\overline{f(p)}}&=&\int^{\Lambda} d^4k\,I'_0(k,p,\theta)
=\int^{\Lambda} d^4k\,I'_0(k',p,\theta)     \nonumber \\
&=&\int^{\Lambda}d^4k \left\{
(1-d_3)\,I_0(k,p,\theta)+d_3\,I_0(k',p,\theta)\right\} \nonumber \\
&=&(1-d_3)\,f(p)+d_3\,f'(p)\quad.
\label{eq:as5}
\ee
\noindent
Knowing that $k_{\mu}=0$ is the right centre for $I_0$ term in
massless QED and
required to satisfy the
massless limit when $M(p^2)\longrightarrow 0$, we  should also choose the
centre
at $k_{\mu}=0$ for the massive case. This means
${\overline{f(p)}}=f(p)$, then  Eq.~(\ref{eq:as5}) becomes~:

\be
0=d_3\,\left[f'(p)-f(p)\right]\quad,
\ee

\noindent
Due to the fact that [$f'(p) \neq f(p)$], $d_3$ should be~:

\be
d_3=0\quad.
\ee

This is just the prescription used in the modified UV cut-off scheme 
which we have already seen gives such excellent numerical agreement with the 
dimensional regularization studies.

\normalsize

\section{Conclusions and Outlook}
\label{sec_conclusions}

 Studying Quantum Electrodynamics necessarily introduces divergences.
 We have seen explicitly that the violation of translational invariance in
nonperturbative studies using an ultraviolet cut-off leaves an error
in the renormalized result.
Hence, if one wants to use cut-off regularization and find the
translationally invariant answer
for the calculated renormalized quantity then a modification is needed. Since
the nonperturbative quantity must give  the perturbative  result in the
weak coupling limit, it is simplest to attempt to  identify these
modifications within  perturbation theory. Once this
is  done we can attempt to generalise it to the nonperturbative case.
Fortunately,
in the case of the fermion self-energy calculation one can establish
a nonperturbative framework on top of the perturbative one.

In this work the
violation of translational invariance for the electron self-energy
is analyzed in
detail and a prescription is presented in order to calculate the 
quantity without
breaking translational invariance in cut-off regularization.
More precisely we mean that violations of translational invariance
are no worse than logarithmic and so for the subtracted (i.e.,
renormalized) integral in the limit $\Lambda\,\longrightarrow\, \infty$,
translational invariance is restored. In this regard, the electron
self-energy is used a test case and as a result we have
seen that a suitably modified cut-off scheme and the dimensional
regularization scheme should be in agreement. Careful numerical studies
(see Tables \ref{massive_cmp_a_table}--\ref{massless_cmp_table} and
Figs.~\ref{sub25fitins.eps}--\ref{submfitins.eps} and
~\ref{subz25fitins.eps}--\ref{subzfit.eps})
have demonstrated this agreement to high precision.
In closing we note that while
we can always add terms with arbitrary coefficients which would vanish
in a translationally invariant regularization scheme, our approach
will only be useful when there are sufficient known constraints to determine
these coefficients uniquely. We are currently attemping to extend this
approach
to include the photon self energy, so that we can study unquenched QED$_4$ 
using a translationally invariant ultraviolet cut-off regularization scheme.

\begin{acknowledgements}

We thank A.~Schreiber for numerious helpful discussions. AGW also 
acknowledges support from the Department of Energy Contract 
No.~DE-FG05-86ER40273 and by the Florida State University Supercomputations 
Research Institute which is partially funded by the Department of Energy 
through contract No.~DE-FC05-85ER25000  
\end{acknowledgements}

%=======================================================================
%          Appendices:
%-----------------------------------------------------------------------
%%%%%%%\begin{appendix}

%%%%%%%\section{Title of appendix goes here}
%%%%%%%\label{appdx_a}

%%%%%%%\end{appendix}

%=======================================================================
%          Bibliography:
%-----------------------------------------------------------------------

\newpage
%=======================================================================
%        Tables:
%-----------------------------------------------------------------------
\begin{table}
  \caption{Absolute percentage comparison of finite renormalization
    $A(p^2)$ for $\epsilon \to 0$ with modified and naive UV cut-off
    massive solutions, with parameters as in Fig.~\ref{sub25eps.eps}. 
    Here $\nu$ represents the scale of the
    extrapolated solutions, while `degree' is the degree of the
    polynomial used to fit the extrapolated solution at each point.
    Higher degree polynomials show no variation from the quintic.}
  \begin{tabular}{cccllllll}
  $\nu$ & degree  & cutoff & $p^2 = 10^{-6}$   & $p^2 = 10^{-2}$
        & $p^2 = 10^{2} $    & $p^2 = 10^{6} $
        & $p^2 = 10^{10}$    & $p^2 = 10^{14}$      \\
  \hline
   1000 & 4 & mod  	& $2.8\times10^{-5}$ & $2.8\times10^{-5}$
            		& $2.9\times10^{-5}$ & $2.2\times10^{-5}$
            		& $2.3\times10^{-5}$ & $4.7\times10^{-5}$   \\

        &   & naive	& $3.7\times10^{-2}$ & $3.7\times10^{-2}$
        	    	& $3.7\times10^{-2}$ & $3.2\times10^{-2}$
            		& $3.3\times10^{-2}$ & $9.8\times10^{-2}$ \\
  \hline
   1000 & 5 & mod  	& $2.8\times10^{-5}$ & $2.8\times10^{-5}$
	            	& $2.8\times10^{-5}$ & $2.2\times10^{-5}$
        	    	& $2.2\times10^{-5}$ & $5.8\times10^{-5}$ \\

        &   & naive	& $3.7\times10^{-2}$ & $3.7\times10^{-2}$
        		& $3.7\times10^{-2}$ & $3.2\times10^{-2}$
        		& $3.3\times10^{-2}$ & $9.8\times10^{-2}$ \\
  \hline
  10000 & 4 & mod	& $2.6\times10^{-5}$ & $2.6\times10^{-5}$
		        & $2.8\times10^{-5}$ & $2.9\times10^{-5}$
			& $1.2\times10^{-5}$ & $1.5\times10^{-5}$ \\

        &   & naive	& $3.7\times10^{-2}$ & $3.7\times10^{-2}$
        		& $3.7\times10^{-2}$ & $3.2\times10^{-2}$
        		& $3.3\times10^{-2}$ & $9.8\times10^{-2}$ \\
  \hline
  10000 & 5 & mod	& $2.5\times10^{-5}$ & $2.5\times10^{-5}$
		     	& $2.8\times10^{-5}$ & $3.0\times10^{-5}$
		     	& $1.2\times10^{-5}$ & $1.8\times10^{-5}$ \\

        &   & naive	& $3.7\times10^{-2}$ & $3.7\times10^{-2}$
        		& $3.7\times10^{-2}$ & $3.2\times10^{-2}$
        		& $3.3\times10^{-2}$ & $9.8\times10^{-2}$ \\
  \end{tabular}
  \label{massive_cmp_a_table}
\end{table}

\begin{table}
  \caption{Absolute percentage comparison as above, but for the mass
    function $M(p^2)$.}
  \begin{tabular}{cccllllll}
  $\nu$ & degree & cutoff & $p^2 = 10^{-6}$    & $p^2 = 10^{-2}$
        & $p^2 = 10^{2} $    & $p^2 = 10^{6} $
        & $p^2 = 10^{10}$    & $p^2 = 10^{14}$  \\
  \hline
    1000 & 4 & mod	& $0.00012$ & $0.00012$
		        & $0.00012$ & $0.00013$
		        & $0.00026$ & $0.055$   \\
         &   & naive	& $0.33$    & $0.33$
		        & $0.33$    & $0.33$
		        & $0.63$    & $1.9$   \\
  \hline
    1000 & 5 & mod	& $0.00012$ & $0.00012$
		        & $0.00012$ & $0.00013$
		        & $0.00026$ & $0.054$   \\
         &   & naive	& $0.33$    & $0.33$
		        & $0.33$    & $0.33$
		        & $0.63$    & $1.9$   \\
  \hline
   10000 & 4 & mod	& $0.00026$ & $0.00026$
		        & $0.00026$ & $0.00046$
		        & $0.00049$ & $0.035$   \\
         &   & naive	& $0.33$    & $0.33$
		        & $0.33$    & $0.32$
		        & $0.63$    & $2.0$   \\
  \hline
   10000 & 5 & mod	& $0.00026$ & $0.00026$
	        	& $0.00026$ & $0.00046$
	        	& $0.00047$ & $0.035$   \\
         &   & naive	& $0.33$    & $0.33$
		        & $0.33$    & $0.32$
		        & $0.63$    & $2.0$   \\
  \end{tabular}
  \label{massive_cmp_m_table}
\end{table}

\begin{table}
  \caption{Absolute percentage comparison of the finite renormalization
    $A(p^2)$ for $\epsilon \to 0$ with modified and naive UV cut-off
    massless solutions, with parameters as in 
    Fig.~\ref{subz25epscloser.eps}. Here `degree' is the degree of the
    polynomial used to fit the extrapolated solution at each point.
    Higher degree polynomials show no variation from the quintic.}
  \begin{tabular}{ccllllllc}
    degree  & cutoff	& $p^2 = 10^{-10}$  & $p^2 = 10^{-5}$
		   	& $p^2 = 10^{0}$    & $p^2 = 10^{5}$
        	   	& $p^2 = 10^{10}$   & $p^2 = 10^{15}$
	             	& $p^2 = 10^{20}$   \\
  \hline
    4   & mod		& $6.9\times10^{-5}$    & $6.2\times10^{-6}$
		        & $2.1\times10^{-6}$    & $5.3\times10^{-7}$
		        & $7.3\times10^{-6}$    & $1.4\times10^{-4}$
		        & $8.4\times10^{-4}$    \\
        & naive		& $2.9\times10^{-1}$	& $2.1\times10^{-1}$
		        & $1.3\times10^{-1}$	& $4.9\times10^{-2}$
		        & $3.3\times10^{-2}$	& $1.1\times10^{-1}$
		        & $1.9\times10^{-1}$  \\
  \hline
    5   & mod		& $6.8\times10^{-5}$    & $6.1\times10^{-6}$
		        & $2.1\times10^{-6}$    & $5.8\times10^{-7}$
		        & $7.5\times10^{-6}$    & $1.4\times10^{-4}$
		        & $8.5\times10^{-4}$    \\
        & naive		& $2.9\times10^{-1}$	& $2.1\times10^{-1}$
		        & $1.3\times10^{-1}$	& $4.9\times10^{-2}$
		        & $3.3\times10^{-2}$	& $1.1\times10^{-1}$
		        & $1.9\times10^{-1}$  \\
  \end{tabular}
  \label{massless_cmp_table}
\end{table}

\newpage
%=======================================================================
%        Figures:
%-----------------------------------------------------------------------

\begin{figure}[htb]
  \setlength{\epsfxsize}{20.0cm}
  \centering
      \epsffile{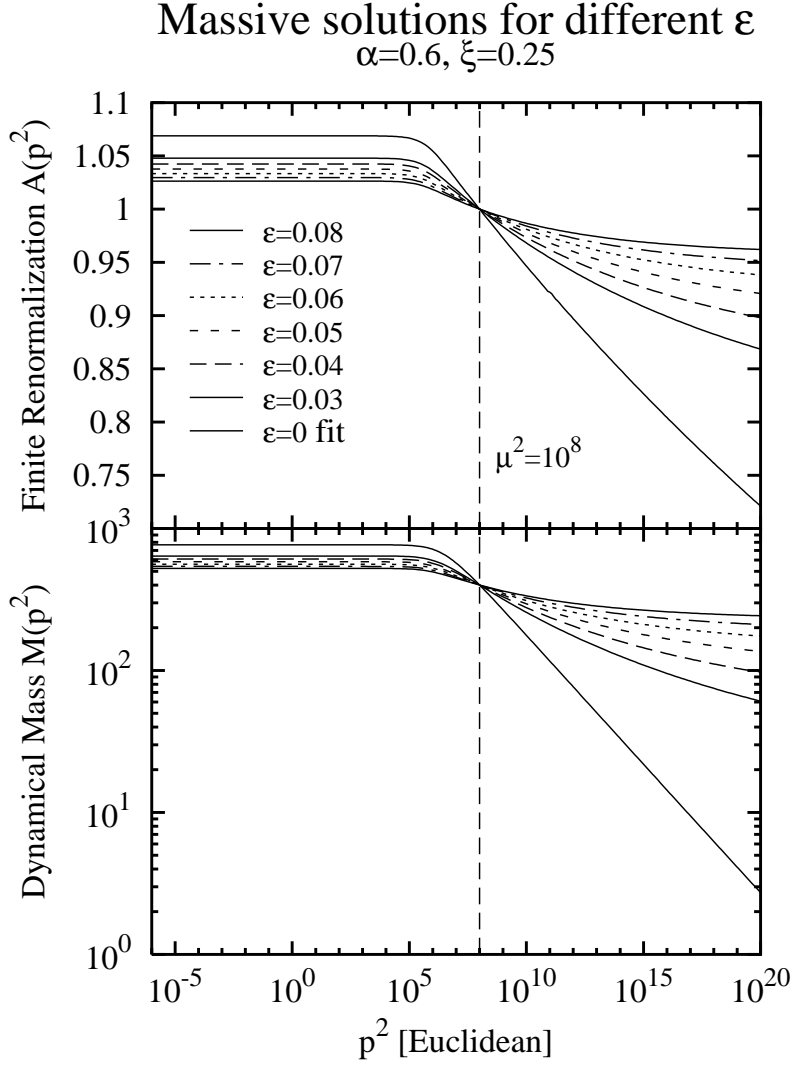}
  
\vspace{10mm}

  \parbox{120mm}{\caption{
    The finite renormalization $A(p^2)$ and mass function $M(p^2)$
    of massive solutions of the fermion DSE for various choices of
    the regulator parameter $\epsilon$, with coupling $\alpha=0.6$,
    gauge parameter $\xi=0.25$, renormalization point $\mu^2 = 10^8$,
    renormalized mass $m(\mu) = 400$ and scale $\nu=1$.
    Also shown are $A(p^2)$ and $M(p^2)$ extrapolated to $\epsilon=0$,
    obtained by fitting polynomials quartic in $\epsilon$ at each
    momentum point.
  \label{sub25eps.eps}}}
\end{figure}

\begin{figure}[htb]
  \vspace{0.5cm}
  \setlength{\epsfxsize}{20.0cm}
  \centering
      \epsffile{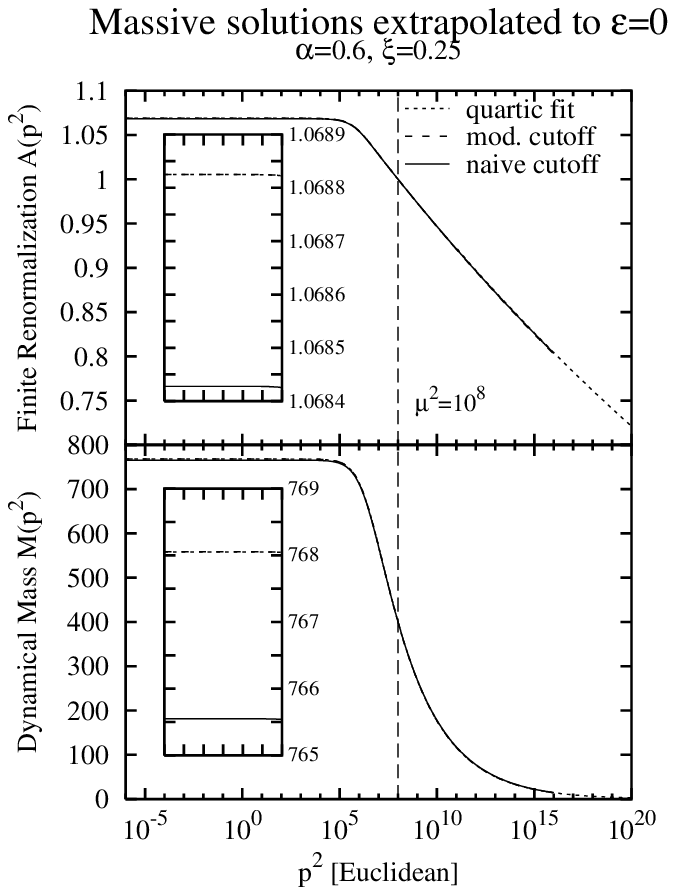}

\vspace{1cm}

  \parbox{120mm}{\caption{
    The finite renormalization $A(p^2)$ and mass function $M(p^2)$
    of the massive solution of the fermion DSE extrapolated to
    $\epsilon=0$ from Fig.~\ref{sub25eps.eps}, compared with
    solutions using naive and modified UV cut-off regulators.
    Note that the curves are almost indistinguishable on the main
    figure: the inserts in the figure for
    the infra-red reveal the excellent agreement between the
    extrapolated (quartic fit) dimensional regularization
    solutions and the modified cut-off solutions which lie
    on top of each other.
  \label{sub25fitins.eps}}}
\end{figure}

\begin{figure}[htb]
  \vspace{0.5cm}
  \setlength{\epsfxsize}{20.0cm}
  \centering
      \epsffile{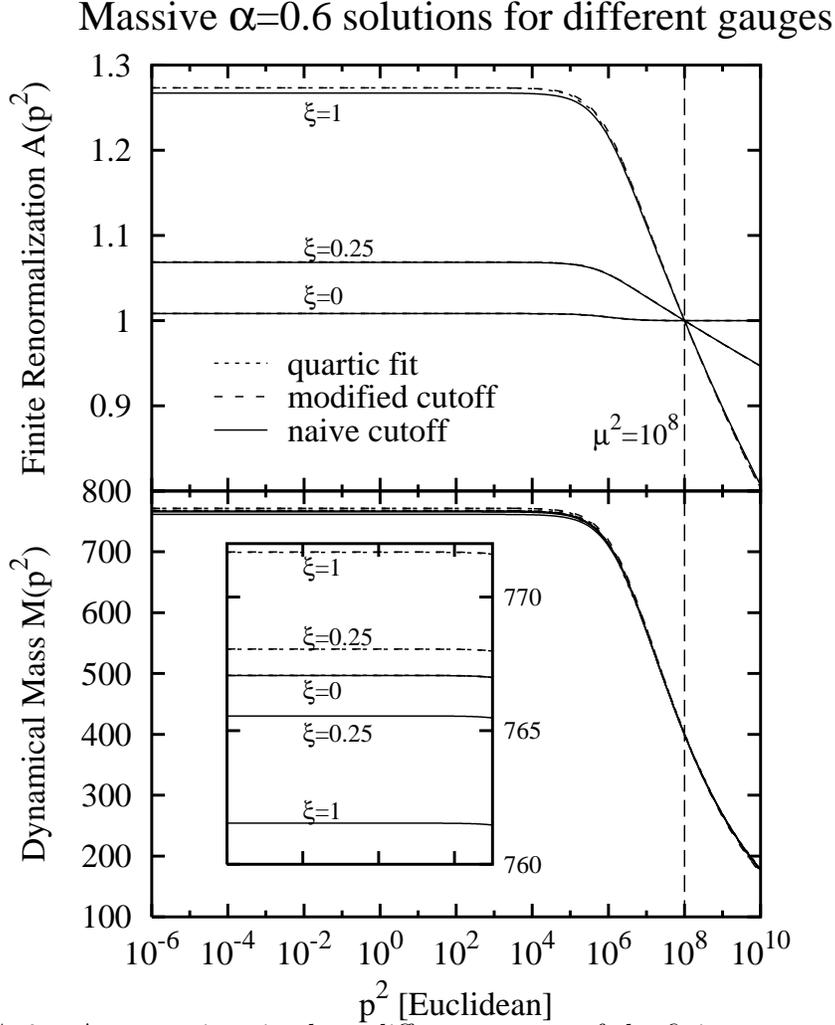}
\vspace{1cm}
  \parbox{120mm}{\caption{
    A comparison in three different gauges of the finite renormalization
    $A(p^2)$ and mass function $M(p^2)$ of the massive solution of the
     fermion DSE extrapolated to $\epsilon=0$, and solutions using
    naive and modified UV cut-off regulators. Other parameters are the
    same as in Fig.~\ref{sub25eps.eps}. As in the massless case,
    the $A(p^2)$ solutions are identical in Landau gauge.  The
    agreement between the extrapolated (quartic) dimensional
    regularization
    solution and the modified cut-off
    solution is readily seen in the Feynman gauge and the curves
    again lie on top of each other in all gauges. 
    The naive cut-off solution clearly disagrees with dimensional
    regularization result.  Owing to the
    approximate gauge invariance of the mass function, an insert is
    neccessary to reveal that the same holds for $M(p^2)$.
  \label{submfitins.eps}}}
\end{figure}

\begin{figure}[htb]
  \setlength{\epsfxsize}{20.0cm}
  \centering
      \epsffile{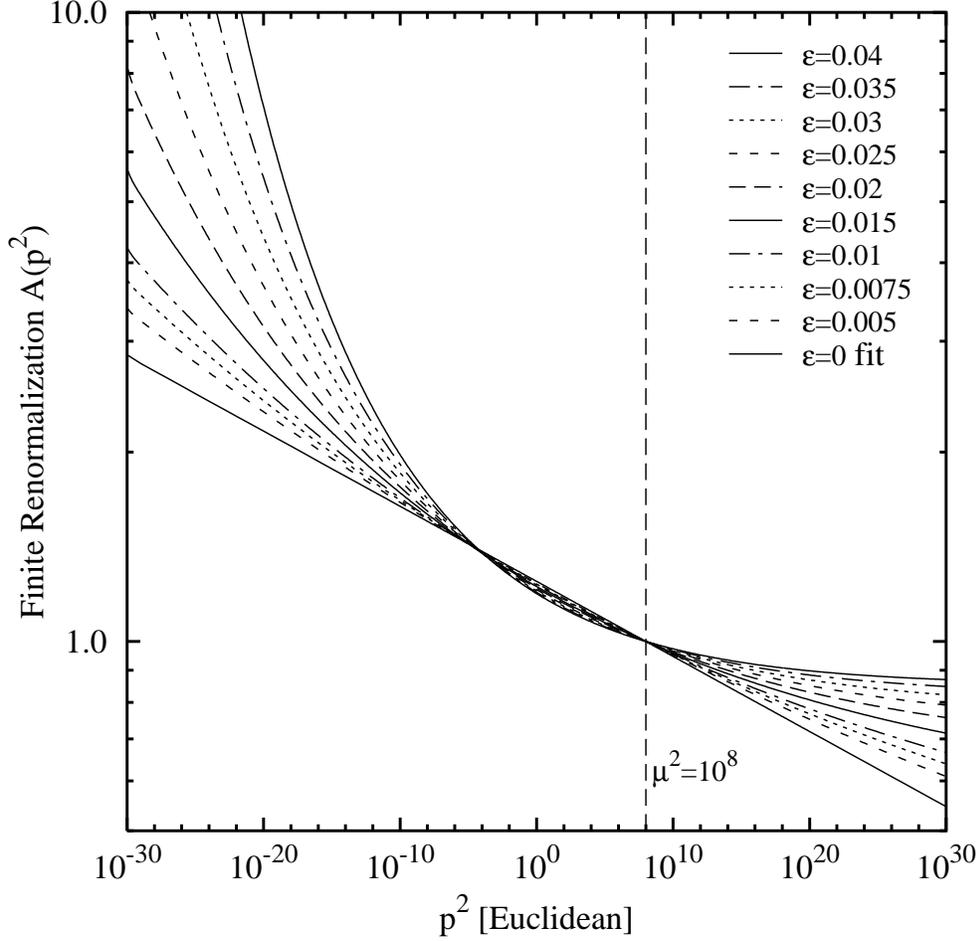}
 
\vspace{1cm}

 \parbox{120mm}{\caption{
    The finite renormalization $A(p^2)$ of chirally symmetric zero-mass
    solutions of the fermion DSE for various choices of the regulator
    parameter $\epsilon$, with coupling $\alpha=0.6$, gauge parameter
    $\xi=0.25$, renormalization point $\mu^2 = 10^8$ and scale $\nu=1$.
    Also shown is $A(p^2)$ extrapolated to $\epsilon=0$, obtained by
    fitting a polynomial quartic in $\epsilon$ to $\log_{10}(A)$
    at each point.
    The logarithmically scaled axes reveals the power-law character of
    the extrapolated solution (straight line is this figure).
    The invariance in $\epsilon$ of the point at $p^2 \approx 10^{-4}$       
    is only approximate, the spread being about 5 parts in $10^3$.
  \label{subz25epscloser.eps}}}
\end{figure}

\begin{figure}[htb]
  \vspace{0.5cm}
  \setlength{\epsfxsize}{20.0cm}
  \centering
      \epsffile{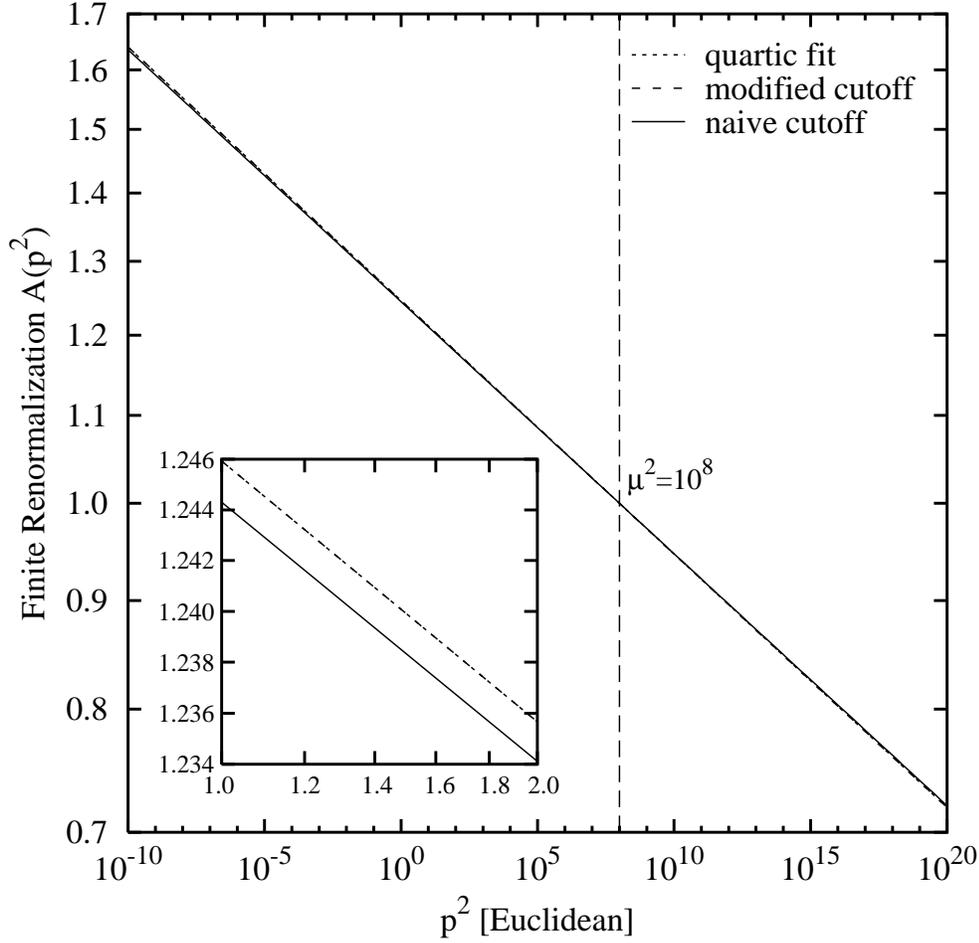}

\vspace{1cm}

  \parbox{120mm}{\caption{
    The finite renormalization $A(p^2)$ of the chirally symmetric
    zero-mass solution of the fermion DSE extrapolated to $\epsilon=0$
    from Fig.~\ref{subz25epscloser.eps},
    compared with the formulae (\ref{massless-naive-cutoff-soln}) and
    (\ref{massless-mod-cutoff-soln}) for the naive
    and modified UV cut-off solutions respectively. The curves are
    indistinguishable on the main figure: an insert reveals the
    extrapolated solution agrees with the modified cut-off solution.
  \label{subz25fitins.eps}}}
\end{figure}

\begin{figure}[htb]
  \vspace{0.5cm}
  \setlength{\epsfxsize}{20.0cm}
  \centering
      \epsffile{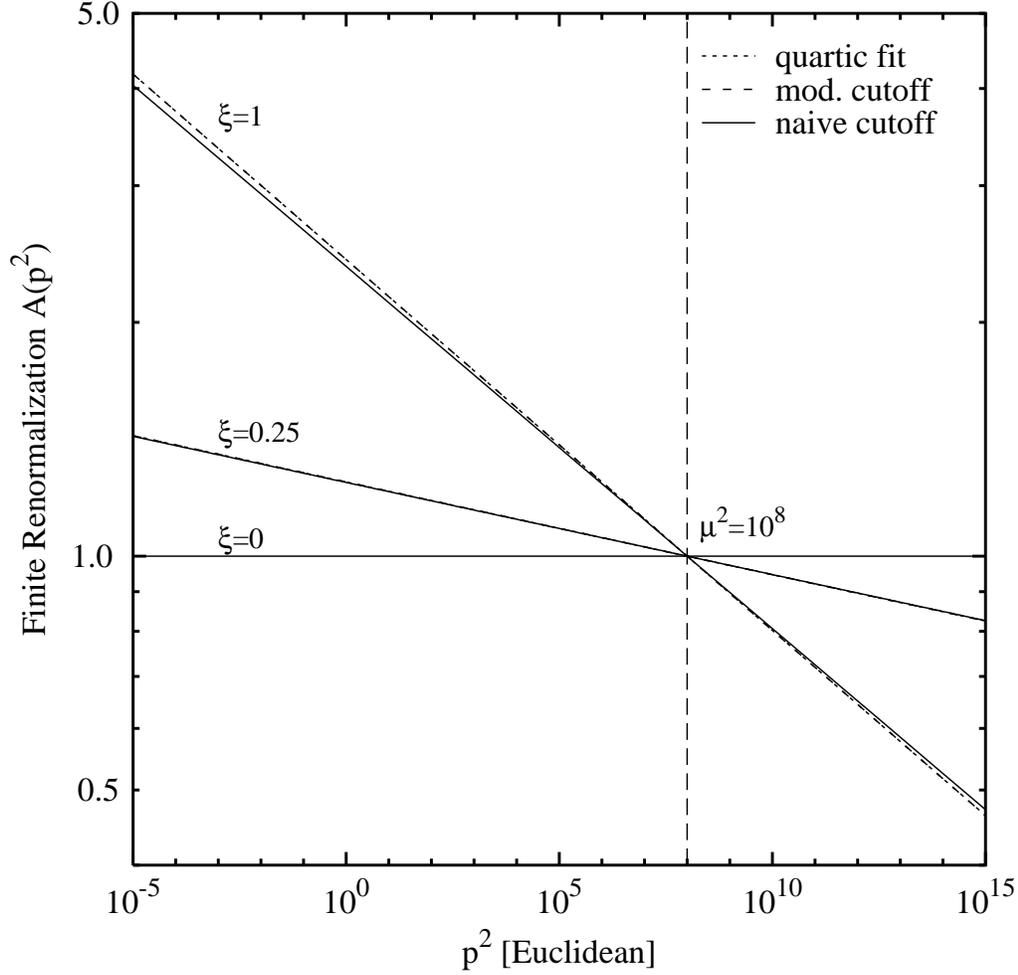}
 
\vspace{1cm}

 \parbox{120mm}{\caption{
    A comparison in three different gauges of the finite renormalization
    $A(p^2)$ of the chirally symmetric zero-mass solution extrapolated to
     $\epsilon=0$ and the naive and modified UV cut-off solutions given by
    eqns. (\ref{massless-naive-cutoff-soln}) and
    (\ref{massless-mod-cutoff-soln}) respectively.
    Other parameters are the same as in Fig.~\ref{subz25epscloser.eps}.
    The curves are identical in Landau gauge: in Feynman gauge, the
    agreement between the extrapolated and modified UV cut-off solutions
    is clearly visible.
  \label{subzfit.eps}}}
  \vspace{0.5cm}
\end{figure}

\end{document}